\documentclass[final,onecolumn,11pt]{IEEEtran}
\ifCLASSINFOpdf
\else
\fi
\newcommand{\F}{\mathbb{F}}

\newcommand {\AAA}{{\mathcal{A}}}

\newcommand {\EE}{{\mathcal{E}}}

\newcommand {\ccc}{{\mathbf{c}}}
\newcommand{\C}{{\mathcal{C}}}
\newcommand{\D}{{\mathcal{D}}}
\newcommand{\SSS}{{\mathcal{S}}}
\newcommand{\BBB}{{\mathcal{B}}}
\newcommand{\PP}{{\mathcal{P}}}

\newcommand{\Ham}{{\mathrm{Ham}}}
\newcommand{\Supp}{\mathrm{Supp}}

\newcommand{\aaa}{{{\mathbf{a}}}}
\newcommand{\bbb}{{{\mathbf{b}}}}

\newcommand{\uuu}{{{\mathbf{u}}}}
\newcommand{\vvv}{{{\mathbf{v}}}}

\newcommand{\wt}{{{\rm{wt}}}}

\newcommand{\rank}{{\rm rank}}
\usepackage{amsmath}
\usepackage{amssymb}
\usepackage{bm}
\usepackage{amssymb}
\usepackage{amsthm}
\usepackage{multirow,booktabs}
\usepackage{cases}
\usepackage{tabularx}
\usepackage{adjustbox}
\usepackage[figuresright]{rotating}
\usepackage{longtable}
\usepackage{booktabs}
\usepackage{multirow}
\usepackage{color}
\usepackage{cite}
\usepackage{multirow,booktabs}
\usepackage{cases}
\usepackage{tabularx}
\usepackage{adjustbox}
\usepackage[figuresright]{rotating}
\usepackage{longtable}
\usepackage{booktabs}
\usepackage{multirow}
\usepackage{color}
\newtheorem{theorem}{Theorem}

\newtheorem{remark}{Remark}
\newtheorem{definition}[theorem]{Definition}
\newtheorem{lemma}[theorem]{Lemma}
\newtheorem{corollary}[theorem]{Corollary}
\newtheorem{example}[theorem]{Example}
\newtheorem{conjecture}[theorem]{Conjecture}

\begin{document}
\title{On  $\ell$-MDS codes and a conjecture on infinite families of $1$-MDS codes  
\thanks{First and second authors were supported by the National Natural Science Foundation of China (Nos.12171134 and U21A20428). 
Third author was supported by Grant TED2021-130358B-I00 funded by MCIN/AEI/10.13039/501100011033 and by the “European Union NextGenerationEU/PRTR”}}

\author{Yang~Li, Shixin~Zhu$^\dag$\thanks{$^\dag$ Corresponding author}
        and~Edgar~Mart\'inez-Moro% <-this % stops a space
\thanks{Yang~Li and  Shixin~Zhu are with the School of Mathematics, Hefei University of Technology, Hefei 230601, China (e-mail: yanglimath@163.com, zhushixinmath@hfut.edu.cn).}% <-this % stops a space
\thanks{Edgar~Mart\'inez-Moro is with the Institute of Mathematics University of Valladolid, Spain (e-mail: Edgar.Martinez@uva.es).}% <-this % stops a space
\thanks{Manuscript received --; revised --}}

% \author{....lastname \thanks{...} \thanks{...}}
% \thanks is no different in this regard, so shield the last} of each \thanks

\markboth{}%
{Shell \MakeLowercase{\textit{et al.}}: Bare Demo of IEEEtran.cls for Journals}



\maketitle

\begin{abstract}
The class of  $\ell$-maximum distance separable ($\ell$-MDS) codes {is a}  generalization of maximum distance separable (MDS) codes {that} has attracted a lot of attention due to  its  
applications in  several areas such as secret sharing schemes, index coding problems, informed source coding problems and combinatorial $t$-designs. 
In this paper, for $\ell=1$, we completely solve a conjecture recently proposed by Heng $et~al.$ (Discrete Mathematics, 346(10): 113538, 2023) 
and obtain infinite families of $1$-MDS codes with general dimensions holding $2$-designs. 
These later codes are also been proved to be optimal locally recoverable codes.  
For general {positive integers} $\ell$ and $\ell'$, we construct new $\ell$-MDS codes from known $\ell'$-MDS codes via some classical propagation rules 
involving the extended, expurgated, and $(\uuu,\uuu+\vvv)$ constructions. 
Finally, we study some general results including characterization, weight distributions, and bounds on maximum lengths of $\ell$-MDS codes,  
which generalize, simplify, or improve some known results in the literature.

\end{abstract}


% \fi
\IEEEpeerreviewmaketitle

\section{Introduction}\label{sec-introduction}

%\subsection{Background}
Throughout this paper, $\F_q$ denotes the finite field with size $q$ where $q=p^h$ is a prime power, and $\F_q^*=\F_q\setminus \{0\}$ {its group of units}. 
A {\textbf{linear code}} of length $n$ and dimension $k$, denoted by $[n,k]_q$, is a $k$-dimensional linear subspace of {the  vector space $\F_q^n$. 
If an $[n,k]_q$ linear code $\C$ has minimum distance $d$, we {will denote it as} $\C$ {being} an $[n,k,d]_q$ linear code.

For an $[n,k,d]_q$ linear code $\C$, the well-known Singleton bound  {states that} $d\leq n-k+1$, which yields {the definition of} a non-negative integer $\SSS(\C)=n-k-d+1$, 
namely the {\textbf{Singleton defect}} of $\C$ \cite{D1996-defect}.  
Let $\C^{\perp}$ be the {orthogonal} (or dual) code of $\C$ with respect to a certain inner product. 
The concept of $\ell$-maximum distance separable (MDS) codes was introduced independently by Liao $et~al.$ \cite{LL2014-m-MDS-Liao} and Tong $et~al.$ \cite{TCY2014-m-MDS-Tong} in 2014.  
They call $\C$ an {\textbf{$\ell$-MDS code}} or say that $\C$ has the {\textbf{$\ell$-MDS property}} if $\SSS(\C)=\SSS(\C^{\perp})=\ell$.  
%Moreover, $\C$ is called an $m$-MDS code if $\ell=m$, i.e., $\SSS(\C)=\SSS(\C^{\perp})=m$ for a fixed non-negative integer $m$. 
%A linear code $\C$ is called an $m$-MDS code if both $\C$ and its dual code $\C^{\perp}$ have $m$ Singleton defect from being an MDS code. 
The {class of} $\ell$-MDS codes {has} both theoretical and practical significance. 
On one hand, many well-known linear codes are $\ell$-MDS codes, such as  binary and ternary extended Golay codes, quaternary (extended) quadratic-residual codes,  
$q$-ary Hamming codes, algebraic geometric codes, and twisted generalized Reed-Solomon codes 
(see for example \cite{LL2014-m-MDS-Liao,LingXing,MS1977,TV1991-AG codes,GZ2023,BPN2017,SYLH2022}). 
Since the Singleton bound is rough for an $[n,k]_q$ linear code if $n$ is large with respect to $q$, 
$\ell$-MDS codes approximate maximal minimum distance for given $n$, $k$, and small $\ell$.  
The interested reader can refer to \cite{LL2014-m-MDS-Liao,TCY2014-m-MDS-Tong,SYLH2022,GZ2023} for more examples and details. 
On the other hand,  $\ell$-MDS codes also have important applications in secret sharing schemes \cite{MSS2019}, 
binary index coding problems \cite{TR2017} and informed source coding problems \cite{TR2018} as well as combinatorial designs \cite{DT2022}.

In this paper, we focus on the constructions and general results on $\ell$-MDS codes. We have three main motivations and objectives. 
\begin{description}
	
	\item{\bf Motivation 1.-}  {The family of} $1$-MDS codes {provides codes that} have good parameters and are closely related to {some objects in} combinatorial design theory. 
	The first $1$-MDS code, the $[11,6,5]_3$ Golay code, was discovered by Golay in 1949, and   
	it holds $4$-designs. %and its extended code holds a Steiner system $S(5,6,12)$ with the largest strength known. 
	Seventy years later, Ding $et~al.$ constructed two infinite families of $1$-MDS codes respectively holding $2$-designs and $3$-designs via cyclic codes \cite{DT2020}. 
	These are the first two infinite families of $1$-MDS codes found that can support designs. 
	Immediately after that, there were found some infinite families of $1$-MDS codes supporting $t$-designs (see, e.g., \cite{TD2020,HW2023Conj,XCQ2022,YZ2022}). 
	It must be noticed that Heng $et~al.$ proposed in  \cite{HW2023Conj} a conjecture on infinite families of $1$-MDS $[q-1,k,q-k-1]_q$ codes holding $2$-designs for each $3\leq k\leq q-2$ 
	(see Conjecture \ref{conj.1} in  {this paper}). 
	As they {mention in that paper}, the problem of constructing infinite families of $1$-MDS codes with general dimensions holding $t$-designs can be tackled if Conjecture \ref{conj.1} is true. 
	
	Thus, {\bf the first objective} of this work is to solve Conjecture \ref{conj.1}. 
	The main results {towards tackling with} this objective can be summarized as follows: 
	\begin{itemize}
		\item Based on a very recent work of Han $et~al.$ \cite{HZ2023}, Conjecture \ref{conj.1} is  {disproved in the cases} $k=q-2$ or $q-3$ in Theorem \ref{th.conj false}. 
		
		\item We then employ subset sum theory to prove that Conjecture \ref{conj.1} is indeed true for left cases in Theorems \ref{th.conj111} and \ref{th.conj333 weight distribution}.  
		Moreover, we also prove that these infinite families of $1$-MDS codes  {provide} optimal locally recoverable codes in Theorem~\ref{th.optimal LRC}. 
	\end{itemize}
	%We first show that Conjecture \ref{conj.1} is invalid if $k=q-2$ or $q-3$. 
	%Then we employ subset sum theory to prove that Conjecture \ref{conj.1} is true for each $3\leq k\leq q-4$. 
	%In summary, Conjecture \ref{conj.1} proposed by Heng $et~al.$ has been completely solved. 

	\item{\textbf{Motivation 2.-}}  Tong $et~al.$  {in \cite{TCY2014-m-MDS-Tong}} employed  punctured and shortened codes of a known $\ell$-MDS code 
	to derive new $\ell$-MDS codes under some conditions (see Lemma \ref{lem.m-MDS codes via PS codes} in this paper).   
	%Note that the new codes obtained from Lemma \ref{lem.m-MDS codes via PS codes} have the same $\ell$-MDS property with the initial code. 
	Recall that, for the index coding problem studied in \cite{TR2017} and the informed sourcing coding problem studied in \cite{TR2018}, 
	one would prefer to get $\ell$-MDS codes with large $\ell$. 
	Hence, based on a known $\ell$-MDS code, if one can obtain a new $\ell'$-MDS code with $\ell'> \ell$, then the new code 
	may be able to provide solutions for an index coding problem or an informed sourcing coding problem even if the original one can not. 
	In addition, we also note that sufficient and necessary conditions for $q$-ary twisted generalized Reed-Solomon codes with length $n\leq q$ 
	to be $\ell$-MDS have been characterized for general $\ell$ by Gu $et~al.$ in \cite{GZ2023}. 
	%As a consequence, a natural and meaningful problem arises as follows: 
	%\begin{problem}\label{prob.2}
	%    Can new $m'$-MDS codes be constructed from known $m$-MDS codes satisfying $m'\neq m$? 
	%\end{problem}
	
	{Based on the previous discussion, our} \textbf{second objective} will be to obtain $q$-ary $\ell$-MDS codes with length $n>q$ and large $\ell$. 
	We study $\ell$-MDS properties of some classical propagation rules.  {The} main results on this objective can be summarized as follows: 
	\begin{itemize}
		\item Based on a known binary $\ell_1$-MDS code, we study the $\ell$-properties of its extended code and expurgated code in 
		Theorems \ref{th.m-MDS code from extended codes} and \ref{th.m-MDS code from expurgated codes}, respectively. 
		Optimal binary $\ell$-MDS codes can be derived in these two ways and we present some concrete examples in Examples \ref{exam.extended Hamming codes} and \ref{exam.m-MDS codes from extended codes}. 
		
		\item Based on two known $q$-ary $\ell_1$-MDS and $\ell_2$-MDS codes, we study the $\ell$-MDS property of their $(\uuu,\uuu+\vvv)$ construction in Theorem \ref{th.m-MDS codes from (u,u+v) construction}.  
		Some new infinite families of $q$-ary $\ell$-MDS codes with length $n>q$ can be obtained in this way.  
	\end{itemize}

	\item{\textbf{Motivation 3.-}} Note that some properties on $1$-MDS, $2$-MDS and $3$-MDS codes have been specifically studied in 
	\cite{LL2014-m-MDS-Liao,DL1995-NMDS,T2012-NNMDS,TCY2014-m-MDS-Tong}. 
	However, it seems that there are little general results on $\ell$-MDS codes in the literature.  
	Due to important applications of $\ell$-MDS codes, it is natural to study some general results of $\ell$-MDS codes. 
	
	{Therefore, our \textbf{third objective} is to} study the characterization, weight distributions, and bounds on maximum lengths of $\ell$-MDS codes in 
	Theorems \ref{th.d(C)+d(DC)=n-2m+2}, \ref{th.weight distribution}, \ref{th.m-mds bounds}, and \ref{th.bounds for q=2}. 
	These results generalize, simplify, or improve the known ones in the literature. 
\end{description}

Based on these motivations this paper is organized as follows. After the introduction, Section~\ref{sec2} reviews some previous basic and useful  {notions and results to keep the paper as  self-contained as possible}. 
Section~\ref{sec3} discusses a conjecture on $1$-MDS codes in detail.  
As a result, we obtain infinite families of $1$-MDS codes holding $2$-designs and optimal locally recoverable codes. In 
Section~\ref{sec4}, we focus the study  on the $\ell$-MDS properties of some classical propagations including extended codes, 
expurgated codes as well as the $(\uuu,\uuu+\vvv)$ construction. 
Section~\ref{sec5} studies some general results of $\ell$-MDS codes. 
Finally, in Section~\ref{sec6},  {we point some concluding remarks on the topic}.

\section{Preliminaries}\label{sec2}

As stated before,  $\F_q$ denotes the finite field with size $q$ where $q=p^h$ is a prime power, and $\F_q^*=\F_q\setminus \{0\}$ {its group of units}. 
Let $\C$ be a linear code over $\mathbb F_q$.  
A vector  $\mathbf c\in \C$ will be called a codeword of $\C$ and the   weight of $\mathbf c$ is the number of non-zero coordinates in it.
For {an integer} $0\leq i\leq n$, let $A_i$ be the number of codewords with weight $i$ in $\C$.  {We will}  denote by $A(z)=1+A_1z+A_2z^2+\cdots+A_nz^n$ the {\textbf{polynomial weight enumerator}} of $\C$.  {The sequence}
$\{A_i\mid i=0,1,\ldots,n\}$ {is called} the {\textbf{weight distribution}} of $\C$, which contains crucial information on the error correction capabilities of this code and 
allows one to compute the error probability of its error correction \cite{K2007-error correction}.   {A square matrix} $M$  {with entries in $\mathbb F_q$}, and with exactly one nonzero entry in each row and each column, and all zero entries in other places, is called a a {\textbf{monomial matrix}}. 
Two linear codes $\C_1$ and $\C_2$ {are} said to be {\textbf{monomially equivalent}}, if there exists a monomial matrix $M$ such that $\C_1=\C_2\cdot M$, {where $\C_2\cdot M$ is the set whose elements are $\mathbf c\cdot M$, $\mathbf c\in \C_2$}. {It is easy to check that}
two monomially equivalent linear codes have the same length, dimension, minimum distance, and weight distribution.  

\subsection{Galois $\ell$-MDS codes}

For any two vectors $\mathbf{x}=(x_1,x_2,\ldots,x_n)$ and $\mathbf{y}=(y_1,y_2,\ldots,y_n)\in \F_q^n$, 
  their {\textbf{$e$-Galois inner product}} is defined as 
\begin{equation}
	\langle \mathbf{x},\mathbf{y} \rangle_e = \sum_{i=1}^{n}x_iy_i^{p^e},\ {\rm {where}}\ 0\leq e\leq h-1, 
\end{equation}
which is a generalization of the {Euclidean inner product} ({when} $e=0$) and the {Hermitian inner product} (when $e=\frac{h}{2}$ with $h$ an even integer). 
The {\textbf{$e$-Galois orthogonal or dual code}} of $\C$  {is defined as} 
\begin{equation}
	\mathcal{C}^{\perp_e}=\{\mathbf{y}\in \mathbb{F}_{q}^n\mid \langle \mathbf{x},\mathbf{y} \rangle_e=0\ {\rm {for\ all}}\ \mathbf{x}\in \mathcal{C}\}.
\end{equation}
Hence, $\C^{\perp_0}$ (resp. $\C^{\perp_{\frac{h}{2}}}$  {if} $h$  {is even}) is the Euclidean (resp. Hermitian) {dual code} of $\C$. 
Let $\sigma: \mathbb{F}_q\rightarrow \mathbb{F}_q$, $a\mapsto a^p$ 
be the {\textbf{Frobenius automorphism}} of $\mathbb{F}_q$. 
For any $[n,k,d]_q$ linear code $\C\subseteq \mathbb F_q^n$,  {we extend it coordinate-wise, that is},
$
	\sigma(\C)=\{\sigma(\ccc)=(\sigma(c_1),\sigma(c_2),\ldots,\sigma(c_n))\mid \ccc=(c_1,c_2,\ldots,c_n)\in \C)\}.
$  
The following result is well-known. 
\begin{lemma}[Lemma 2.3 in \cite{Galois LCD} and Proposition 2.2 in \cite{Panxu}]\label{lem.Galois proposition}
	Let $q=p^h$ and $\C$ be an $[n,k,d]_q$ linear code. 
	Then for any $0\leq e\leq h-1$, the following statements hold. 
	\begin{enumerate}
		\item  $\mathcal{C}^{\perp_e}=(\sigma^{h-e}(\mathcal{C}))^{\perp_0}=\sigma^{h-e}(\mathcal{C}^{\perp_0})$. 
		
		\item   $\sigma^{h-e}(\mathcal{C})$ is also an $[n,k,d]_q$ linear code. 
	\end{enumerate}
\end{lemma}
%Liu $et~al.$ \cite{Galois LCD} proved that $\sigma^{h-e}(\mathcal{C})$ is also an $[n,k,d]_q$ linear code and 
%Liu $et~al.$ \cite{Panxu} proved that $\mathcal{C}^{\perp_e}=(\sigma^{h-e}(\mathcal{C}))^{\perp_0}=\sigma^{h-e}(\mathcal{C}^{\perp_0})$. 

%In \cite{D1996-defect}, De Boer defined the {{Singleton defect}} of an $[n,k,d]_q$ linear code $\C$ as $\SSS(\C)=n-k+1-d.$  
To evaluate the parameters of a linear code with $e$-Galois hulls of arbitrary dimensions, 
Li $et~al.$ \cite{LZDM} generalized the concept of Euclidean $\ell$-MDS codes proposed originally by Liao $et~al.$ \cite{LL2014-m-MDS-Liao} and Tong $et~al.$ \cite{TCY2014-m-MDS-Tong} to 
general $e$-Galois $\ell$-MDS codes in the following way. 

\begin{definition}\label{def.m-MDS}
	An $[n,k,d]_q$ linear code is called an {\textbf{$e$-Galois $\ell$-MDS code}} or said to have the {\textbf{$e$-Galois $\ell$-MDS property}} 
	if $\SSS(\C)=\SSS(\C^{\perp_e})=\ell$, where $q=p^h$ and $0\leq e\leq h-1$. 
	%Moreover, if $\SSS(\C)=\SSS(\C^{\perp_e})=m$, $\C$ is specifically called an $e$-Galois $m$-MDS code. 
\end{definition}

Note that Definition \ref{def.m-MDS} is the same  {as} the original definition of Euclidean $\ell$-MDS codes  {when} $e=0$. 
Through the following lemma, we can establish an equivalent relationship between Euclidean $\ell$-MDS codes and $e$-Galois $\ell$-MDS codes.  

%Since $\C^{\perp_e}=(\C^{p^{h-e}})^{\perp_0}=(\C^{\perp_0})^{p^{h-e}}$, then $\C^{\perp_e}$ and $\C^{\perp_0}$ have the same parameters. Hence, $\C$ is Euclidean $m$-MDS if and only if $\C$ is $e$-Galois $m$-MDS. Furthermore, we have the following lemma. 

\begin{lemma}\label{prop.m-MDS}
	Let $q=p^h$ and $0\leq e,\ e'\leq h-1$ be integers. 
	Let $\C$ be an arbitrary linear code. Then $\C$ is $e$-Galois $\ell$-MDS 
	if and only if $\C$ is $e'$-Galois $\ell$-MDS. 
\end{lemma} 
\begin{proof}
	%With Definition \ref{def.m-MDS}, $\C$ is $e$-Galois $\ell$-MDS if and only if $\SSS(\C)=\SSS(\C^{\perp_e})$ 
	%and  $\C$ is $e'$-Galois $\ell$-MDS if and only if $\SSS(\C)=\SSS(\C^{\perp_{e'}})$. 
	It follows from Lemma \ref{lem.Galois proposition} 1) that $\C^{\perp_e}=\sigma^{h-e}(\C^{\perp_0})$ 
	and $\C^{\perp_{e'}}=\sigma^{h-e'}(\C^{\perp_0})$. 
	With Lemma \ref{lem.Galois proposition} 2), we know that both $\C^{\perp_e}$ and $\C^{\perp_{e'}}$ 
	have the same parameters as $\C^{\perp_0}$, which implies that $\SSS(\C^{\perp_e})=\SSS(\C^{\perp_{e'}})$. 
	Hence,  we get the result. 
\end{proof}

Lemma~\ref{prop.m-MDS}  {allows us to only} consider Euclidean $\ell$-MDS codes in the sequel and abbreviate them as $\ell$-MDS codes. 
Moreover, we also  denote $\C^{\perp_0}$ just as $\C^{\perp}$. 
The following result provides a sufficient condition for a linear code being $\ell$-MDS and the result implies that 
it is generally more difficult to obtain $\ell$-MDS codes as $\ell$ increases. 

\begin{lemma}[Theorem 3.6 in \cite{LL2014-m-MDS-Liao} and Theorem 7 in \cite{TCY2014-m-MDS-Tong}]
	\label{lem.m-MDS sufficient condition}
	If $n>\ell q+\ell+k-1$ and $k>(\ell-1)q+\ell-2$, 
	then every $[n,k,n-k-\ell+1]_q$ code is an $\ell$-MDS code. 
\end{lemma}

For $0\leq i\leq n$, let $A_i$ (resp. $A^{\perp}_i$) be the numbers of codewords  {of} weight $i$ in $\C$ (resp. $\C^{\perp}$). 
Let $\{A_i\mid i=0,1,\ldots,n\}$ (resp. $\{A^{\perp}_i\mid i=0,1,\ldots,n\}$) denote the weight distribution of $\C$ (resp. $\C^{\perp}$). 
In \cite{DL1995-NMDS}, Dodunekov $et~al.$ determined the weight distributions of a $1$-MDS code and its dual code. 
\begin{lemma}[Corollary 4.2 in\cite{DL1995-NMDS}]\label{lem.NMDS weight distribution}
	Let $\C$ be a $1$-MDS $[n,k]_q$ code. If $s\in \{1,2,\ldots,k\}$, then 
	$$A_{n-k+s}=\binom{n}{k-s}\sum_{i=0}^{s-1}(-1)^i\binom{n-k+s}{i}(q^{s-i}-1)+(-1)^s\binom{k}{s}A_{n-k}.$$ 
	If $s\in \{1,2,\ldots,n-k\}$, then 
	$$A^{\perp}_{k+s}=\binom{n}{k+s}\sum_{i=0}^{s-1}(-1)^i\binom{k+s}{i}(q^{s-i}-1)+(-1)^s\binom{n-k}{s}A^{\perp}_{k}.$$ 
\end{lemma}

\begin{lemma}[Theorem 10 in \cite{TCY2014-m-MDS-Tong}]\label{lem.m-MDS codes via PS codes}
	Let $\C$ be an $\ell$-MDS $[n,k]_q$  code. If $n>(\ell-1)q+k+\ell-2$ and $k>(\ell-1)q+\ell-2$, 
	then the following statements hold. 
	\begin{enumerate}
		\item  There exists an $\ell$-MDS $[n-1,k]_q$  code. 
		
		\item  There exists an $\ell$-MDS $[n-1,k-1]_q$ code. 
	\end{enumerate}
\end{lemma}

\subsection{Subset sum problems} 
Let $\mathcal{F}\subseteq \F_q$ and $b\in \F_q$. 
The  {\textbf{subset sum problem}} over $\mathcal{F}$ is to determine 
if there is a subset $\emptyset \neq \{x_1,x_2,\ldots,x_r\}\subseteq \mathcal{F}$ such that 
\begin{align}\label{eq.subset sum problem def}
	x_1+x_2+\cdots+x_r=b.     
\end{align}
For $b=0$, if there exists a subset $\{x_1,x_2,\ldots,x_r\}\subseteq \mathcal{F}$ such that Equation~(\ref{eq.subset sum problem def}) holds, 
we  {say that} $\mathcal{F}$ contains a {\textbf{zero-sum subset of size $r$}}; 
and if there is no zero-sum subset of size $r$ in $\mathcal{F}$,  {say that} $\mathcal{F}$  is {\textbf{$r$-zero-sum free}}. 
Generally, the subset sum problem is known to be NP-complete. 
Let $N(r,b,\mathcal{F})$ be the number of subsets $\{x_1,x_2,\ldots,x_r\}\subseteq \mathcal{F}$ such that Equation (\ref{eq.subset sum problem def}) holds. 
In \cite{LW2008}, Li $et~al.$ determined the exact value of $N(r,b,\mathcal{F})$ for $\mathcal{F}=\F_q^*$.  

\begin{lemma}[Theorem 1.2 in \cite{LW2008}]\label{lem.subset sum}
	Let notations be the same as above. Then 
	\begin{align}\label{eq.subset sum}
		%N(r,b,\F_q^*)=\frac{1}{q}\binom{q-1}{r}+(-1)^{k+\lfloor \frac{r}{p} \rfloor}\frac{v(b)}{q}\binom{\frac{q}{p}-1}{\lfloor \frac{r}{p} \rfloor}, 
		N(r,b,\F_q^*)=\frac{1}{q}\left[\binom{q-1}{r}+(-1)^{k+\lfloor \frac{r}{p} \rfloor} v(b) \binom{\frac{q}{p}-1}{\lfloor \frac{r}{p} \rfloor} \right], 
	\end{align}
	where $v(b)=\left\{\begin{array}{lc}
		-1, & {\rm if}~b\neq 0, \\
		q-1, & {\rm if}~b=0.
	\end{array} \right.$
\end{lemma}

\subsection{Combinatorial $t$-designs}

Let $n\geq k\geq t\geq 1$ be three positive integers. 
Let $\PP$ be a set with $|\PP|=n$ and $\BBB$ be a collection of $k$-subsets of $\PP$. 
If each $t$-subset of $\PP$ is contained in exactly $\lambda$ elements of $\BBB$, 
we call $(\PP,\BBB)$ a {\textbf{$t$-$(n,k,\lambda)$ design}} with 
$b=\frac{\lambda\binom{n}{t}}{\binom{k}{t}}$ blocks. 
If $k=t$ or $n$, $t$-$(n,k,\lambda)$ designs are referred to as trivial ones.  
Let $\BBB^{\perp}$  be the set of the complements of  {all} the blocks in $\BBB$. 
If $(\PP,\BBB)$ is a $t$-$(n,k,\lambda)$ design, then $(\PP,\BBB^{\perp})$ is a 
$t$-$(n,n-k,\lambda^{\perp})$ design, where $\lambda^{\perp}=\frac{\lambda\binom{n-t}{k}}{\binom{n-t}{k-t}}$ 
and we call it the {\textbf{complementary design}} of $(\PP,\BBB)$. 

{Many} linear codes induce $t$-designs,  {see for example} \cite{DT2022}. 
Specifically, let $\C$ be an $[n,k]_q$ linear code and $\PP(\C)=\{1,2,\ldots,n\}$. 
For any codeword $\ccc=(c_1,c_2,\ldots,c_n)\in \C$, its {\textbf{support}} is defined by 
$\Supp(\ccc)=\{1\leq i\leq n\mid c_i\neq 0\}$. Let  $\wt(\ccc)$ be the Hamming weight of $\ccc\in \C$.
One can define $\BBB_w(\C)=\frac{S}{q-1}$, where  {$S$ is the multiset} $$S=\{\{\Supp(\ccc)\mid \wt(\ccc)=w~{\rm and}~\ccc\in \C\}\}.$$  
{Thus,} $\frac{S}{q-1}$ is the multiset derived from dividing the multiplicity of each element in $S$ by $q-1$.   
%There is a coding-theoretical construction of $t$-designs as follows \cite{DT2022}.  
%\begin{lemma}[\cite{DT2022}]\label{lem.t-design} 
% Let notations be the same as above. 
If the pair $(\mathcal{P}(\C),  \mathcal{B}_w(\C))$ is a $t$-$(n,w,\lambda)$ design with $b$ blocks for some $0\leq w\leq n$, 
we say that {\textbf{the code $\C$ supports a $t$-design}}, where 
\begin{align}\label{eq.parameters of t-design}
	\lambda=\frac{A_w\binom{w}{t}}{(q-1)\binom{n}{t}}~{\rm and}~ b=\frac{A_w}{q-1}. 
\end{align}
%\end{lemma}

\begin{lemma}[Proposition 14 in \cite{FW1997-sAMDS}]\label{lem.complement design}
	Let $\C$ be a $1$-MDS code. %For any $\ccc=(c_1,c_2,\ldots,c_n)\in \C$, its support is defined by $\Supp(\ccc)=\{1\leq i\leq n\mid c_i\neq 0\}$. 
	Then for any minimum weight codeword $\ccc$ in $\C$, there exists, up to a multiple, a unique minimum weight codeword $\ccc^{\perp}$ in $\C^{\perp}$ 
	satisfying $\Supp(\ccc)\cap \Supp(\ccc^{\perp})=\emptyset$. 
	Moreover, the number of minimum weight codewords in $\C$ and the number of those in $\C^{\perp}$ are equal. 
\end{lemma}

Based on Lemma \ref{lem.complement design} and {the above discussion}, if the minimum weight codewords of a $1$-MDS $[n,k]_q$ code support a $t$-$(n,n-k,\lambda)$ design, 
then the minimum weight codewords of its dual code {also} support a $t$-$(n,k,\lambda^{\perp})$ design,  
where 
\begin{align}\label{eq.complement design}
	\lambda^{\perp}=\frac{\lambda\binom{n-t}{n-k}}{\binom{n-t}{n-k-t}}.    
\end{align}

Finally, we recall some results on generalized Hamming weights from \cite{W1991-GHW}.  
Let $\Supp(\C)$ be the set of coordinate positions where not all codewords in $\C$ are zero. 
For an $[n,k]_q$ linear code $\C$, its {\textbf{$r$-th generalized Hamming weight}} $d_r(\C)$ is defined by 
\begin{align*}
	d_r(\C)=\min\{|\Supp(\D)|\mid \D~{\rm is~an}~[n,r]_q~{\rm subcode~of}~\C\},~ {\rm where}~1\leq r\leq k. 
\end{align*}
%Wei derived the following well-known results \cite{W1991-GHW}. 

\begin{lemma}[\cite{W1991-GHW}]\label{prop.GHW}
	Let $\C$ be an $[n,k]_q$ linear code and $\C^{\perp}$ be its dual code. 
	Then the following statements hold. 
	\begin{enumerate}
		\item   $0<d(\C)=d_1(\C)<d_2(\C)<\ldots<d_k(\C)\leq n.$ 
		\item   $\{d_r(\C)\mid r=1,2,\ldots,k\}\cup \{n-d_r(\C^{\perp})+1\mid r=1,2,\ldots,n-k\}=\{1,2,\ldots,n\}.$
		\item  {\rm ({\textbf{Generalized Singleton bound}})} $d_r(\C)\leq n-k+r$, $r=1,2,\ldots,k.$ 
	\end{enumerate}
\end{lemma}

\section{Infinite families of $1$-MDS codes}\label{sec3} 

\subsection{A conjecture on $1$-MDS codes}
Throughout this section, let $q=2^h$ with $h\geq 3$ and $\alpha$ be a primitive element of $\F_q$. {We will}
denote $\alpha_i=\alpha^i$ for $1\leq i\leq q-1$ and hence, $\F_q^*=\{\alpha_1,\alpha_2,\ldots,\alpha_{q-1}\}$. 
For $3 \leq k\leq q-2$, {we consider the matrix} 
\begin{align}\label{eq.M_k}
	M_{k} =\left[\begin{array}{ccccc}
		1 & 1 & \cdots & 1 & 1  \\ 
		\alpha_1 & \alpha_2 & \cdots & \alpha_{q-2} & \alpha_{q-1}  \\ 
		\alpha_1^2 & \alpha_2^2 & \cdots & \alpha_{q-2}^2 & \alpha_{q-1}^2  \\ 
		\vdots & \vdots & \vdots & \vdots & \vdots \\ 
		\alpha_1^{k-2} & \alpha_2^{k-2} & \cdots & \alpha_{q-2}^{k-2} & \alpha_{q-1}^{k-2}  \\ 
		\alpha_1^{k} & \alpha_2^{k} & \cdots & \alpha_{q-2}^{k} & \alpha_{q-1}^{k}  
	\end{array}\right].   
\end{align}
and let $\C_{k}$ be the $q$-ary linear code generated by {the rows of} $M_{k}$. 
Note that, if we consider the matrix formed by the $k$ columns $\{i_1, i_2, \ldots, i_k\}$ of $M_k$ 
its determinant is $(\alpha_{i_1} + \alpha_{i_2} + \dots + \alpha_{i_k} ) \prod_{1\leq s<t\leq k}(\alpha_{i_t} - \alpha_{i_s} )$, 
which only vanishes if $\alpha_{i_1} + \alpha_{i_2} + \dots + \alpha_{i_k} =0$. 
Thus, the fact of this matrix being non-singular is related to the (zero) subset sum problem.
In \cite{HW2023Conj}, Heng $et~al.$ proposed the following conjecture.

\begin{conjecture}[Conjecture 36 in \cite{HW2023Conj}]\label{conj.1}
	For each $3\leq k\leq q-2$, the linear code $\C_{k}$ is a $1$-MDS $[q-1,k,q-k-1]_q$ code and 
	the minimum weight codewords of both $\C_{k}$ and its dual $\C_{k}^{\perp}$ support $2$-designs. 
\end{conjecture}

In the following, we focus on Conjecture \ref{conj.1}. We first {disprove} Conjecture \ref{conj.1} {for} $k=q-2$ or $k=q-3$ 
and then we prove that Conjecture \ref{conj.1} indeed holds for each $3\leq k\leq q-4$ in Subsection~\ref{subsection:proof}. 
To this end, we need a very recent result proposed by Han $et~al.$ \cite{HZ2023}.

\begin{lemma}[Proposition 2.5 in \cite{HZ2023}]\label{lem.Han}
	{Let $\C_k$ be a linear code generated by {the rows of} $M_{k}$ in Equation~(\ref{eq.M_k})}. Then the following statements hold. 
	\begin{enumerate}
		\item  The linear code $\C_{k}$ is MDS if and only if $\F_q^*$ is $k$-zero-sum free. 
		
		\item   The linear code $\C_{k}$ is $1$-MDS if and only if $\F_q^*$ contains a zero-sum subset of size $k$. 
	\end{enumerate}
\end{lemma}

%First, we prove that $\C_{q-3}$ and $\C_{q-2}$ are MDS codes rather than $1$-MDS codes, which implies that Conjecture \ref{conj.1} does not hold for $k=q-3$ and $q-2$. 

\begin{theorem}\label{th.conj  false}
	The linear code $\C_{q-3}$ is an MDS $[q-1,q-3,3]_q$ code and the linear code $\C_{q-2}$ is an MDS $[q-1,q-2,2]_q$ code. 
	Moreover, both $\C_{q-3}$ and $\C_{q-2}$ only support trivial designs. 
\end{theorem}
\begin{proof}
	From Lemma \ref{lem.subset sum}, we have that 
	\begin{align*}
		N(q-3,0,\F_q^*) & =\frac{1}{q}\left[ \binom{q-1}{q-3}-(q-1)\binom{\frac{q}{2}-1}{\frac{q}{2}-2} \right]=0~{\rm {and}} \\ 
		N(q-2,0,\F_q^*) & =\frac{1}{q}\left[ \binom{q-1}{q-2}-(q-1)\binom{\frac{q}{2}-1}{\frac{q}{2}-1} \right]=0.  
	\end{align*}
Thus, the set $\F_q^*$ is both $(q-3)$-zero sum free and $(q-2)$-zero-sum free. 
	Hence, we get that $\C_{q-3}$ is an MDS $[q-1,q-3,3]_q$ code and $\C_{q-2}$ is an MDS $[q-1,q-2,2]_q$ code from Lemma \ref{lem.Han} 1). 
	From \cite[Chapter 12]{DT2022}, $\C_{q-3}$ and $\C_{q-2}$ support complete designs and thus trivial, which   completes the proof. 
\end{proof}

\subsection{Infinite families of $1$-MDS codes with general dimensions holding $2$-designs}\label{subsection:proof}

\begin{theorem}\label{th.conj111}
	For each $3\leq k\leq q-4$, the linear code $\C_{k}$ is a $1$-MDS $[q-1,k,q-k-1]_q$ code.  
	Moreover, the set of  minimum weight codewords of   $\C_{k}^{\perp}$ support a $2$-$(q-1,k,\lambda_1)$ design  
	and the set of minimum weight codewords of $\C_{k}$ support a $2$-$(q-1,q-k-1,\lambda_2)$ design, where 
	$$\lambda_1=\sum_{i=0}^{k-2}\sum_{j=0}^{k-2-i}(-1)^{i+j}N(i,j)~{\rm and}~\lambda_2=\frac{\lambda_1\binom{q-3}{k}}{\binom{q-3}{k-2}},$$  
	where 
	$N(i,j)=\left\{\begin{array}{ll}
		\frac{1}{q}\left[ \binom{q-1}{k-i-j-2}+(-1)^{k-i-j+\lfloor\frac{k-i-j}{2}\rfloor-1}(q-1)\binom{\frac{q}{2}-1}{\lfloor \frac{k-i-j}{2} \rfloor-1} \right], & {\rm if}~ij~{\rm is~odd}, \\ \\
		\frac{1}{q}\left[ \binom{q-1}{k-i-j-2}+(-1)^{k-i-j+\lfloor\frac{k-i-j}{2}\rfloor}\binom{\frac{q}{2}-1}{\lfloor \frac{k-i-j}{2} \rfloor-1} \right], & {\rm if}~ij~{\rm is~even}.  
	\end{array} \right.$
\end{theorem}
\begin{proof}
	Since $q=2^h\geq 8$ and $3\leq k\leq q-4$, one can easily check  that $\binom{q-1}{k}> (q-1)\binom{\frac{q}{2}-1}{\lfloor \frac{k}{2} \rfloor}$. 
	Then, it  follows from Lemma~\ref{lem.subset sum} and Lemma~\ref{lem.Han} 2) that $\C_k$ is a $1$-MDS code. 
	Let $x_1$ and $x_2$ be two different elements of $\F_q^*$ and let us consider the following subset sum problem 
	\begin{align}\label{eq.subset problem}
		x_3+x_4+\cdots+x_k=x_1+x_2,~{\rm where}~\{x_3,x_4,\ldots,x_k\}\subseteq \F_q\setminus \{0,x_1,x_2\}.  
	\end{align}
	Let $N(k-2,x_1+x_2,\F_q\setminus \{0,x_1,x_2\})$ denote the number of values $\{x_3,x_4,\ldots,x_k\}$ such that Equation (\ref{eq.subset problem}) holds.   
	Since $q=2^h\geq 8$ is even, by the inclusion-exclusion sieving method 
	(i.e., similar arguments to the proofs of \cite[Theorem 1.3]{LW2008} and \cite[Lemma 4.2]{LW2008}), we have 
	\begin{align}\label{eq.subset sum new}
		\begin{split}
			 N\left(k-2,x_1+x_2,\F_q\setminus \{0,x_1,x_2\}\right)  
			&=  \sum_{i=0}^{k-2}(-1)^i N(k-2-i,x_1+(i+1)x_2,\F_q\setminus \{0,x_1\}) \\
			&= \sum_{i=0}^{k-2}(-1)^i \sum_{j=0}^{k-2-i} (-1)^j N(k-2-i-j,(j+1)x_1+(i+1)x_2,\F_q^*) \\
			 &= \sum_{i=0}^{k-2}\sum_{j=0}^{k-2-i} (-1)^{i+j} N(k-2-i-j,(j+1)x_1+(i+1)x_2,\F_q^*).         
		\end{split}
	\end{align} 
	Note that $(j+1)x_1+(i+1)x_2=0$ if and only if both $i$ and $j$ are odd, if and only if $ij$ is odd as $x_1,x_2\in \F_q^*$ and $x_1\neq x_2$. 
	%For other values of $i$ and $j$, $(j+1)x_1+(i+1)x_2\neq 0$.  
	For sort, we will denote $N(k-2-i-j,(j+1)x_1+(i+1)x_2,\F_q^*)$ as $N(i,j)$. Then we have from Lemma \ref{lem.subset sum} that 
	\begin{align}\label{eq.N(i,j)}
		N(i,j)=\left\{\begin{array}{ll}
			\frac{1}{q}\left[ \binom{q-1}{k-i-j-2}+(-1)^{k-i-j+\lfloor\frac{k-i-j}{2}\rfloor-1}(q-1)\binom{\frac{q}{2}-1}{\lfloor \frac{k-i-j}{2} \rfloor-1} \right], & {\rm if}~ij~{\rm is~odd}, \\ \\
			\frac{1}{q}\left[ \binom{q-1}{k-i-j-2}+(-1)^{k-i-j+\lfloor\frac{k-i-j}{2}\rfloor}\binom{\frac{q}{2}-1}{\lfloor \frac{k-i-j}{2} \rfloor-1}\right], & {\rm if}~ij~{\rm is~even}.  
		\end{array} \right.    
	\end{align}
	%Therefore, $N(i,j)$ is known when $i$ and $j$ are fixed. 
	%Moreover, $N(k-2,x_1+x_2,\F_q\setminus \{0,x_1,x_2\})> 0$ since $3\leq k\leq q-4$. 
	
	Since $\C_k$ is $1$-MDS, so does $\C_k^{\perp}$, which implies that $d(\C_{k}^{\perp})=k$. 
	Let $\ccc=(c_1,c_2,\ldots,c_{q-1})\in \C_{k}^{\perp}$ with $\wt(\ccc)=k$ and $\Supp(\ccc)=\{s_1,s_2,\dots,s_k\}$. 
	Hence, $c_{s_t}=u_{s_t}\in \F_q^*$ for $1\leq t\leq k$ and $c_v=0$ for all $v\in \{1,2,\ldots,q-1\}\setminus \{s_1,s_2,\ldots,s_k\}$. 
	Set $x_t=\alpha^{s_t}$ for $1\leq t\leq k$, where $\alpha$ is a primitive element of $\F_q$. 
	Since $\ccc\in \C_{k}^{\perp}$, we have 
	\begin{align}\label{eq.M_{k,k}u=0}
		M_{k,k}\uuu^T=\mathbf{0},
	\end{align}
	%$$M_{k,k}\uuu^T=\mathbf{0},$$ 
	where $\uuu=(u_{s_1}, u_{s_2}, \ldots, u_{s_k})$, $\mathbf{0}$ is a column vector of length $k$ and 
	$$M_{k,k}=\left[\begin{array}{ccccc}
		1 & 1 & \cdots & 1 & 1 \\
		x_1 & x_2 & \cdots & x_{k-1} & x_k \\
		x_1^2 & x_2^2 & \cdots & x_{k-1}^2 & x_k^2 \\ 
		\vdots & \vdots & \vdots & \vdots & \vdots \\
		x_1^{k-2} & x_2^{k-2} & \cdots & x_{k-1}^{k-2} & x_k^{k-2} \\ 
		x_1^{k} & x_2^{k} & \cdots & x_{k-1}^{k} & x_k^{k}
	\end{array}\right].$$ 
	Note that $\rank(M_{k,k})=k-1$ as $\C_k^{\perp}$ is a $1$-MDS code and 
	the first $k-1$ rows and $k-1$ columns of $M_{k,k}$ form a Vandermonde matrix. 
	Then,  the number of solutions  $\{u_{s_1}, u_{s_2}, \ldots, u_{s_k}\}\subseteq (\F_q^*)^k$ 
	of  Equation~(\ref{eq.M_{k,k}u=0})   equals $q-1$. 
	Furthermore, it implies that all codewords of weight $k$ in $\C_{k}^{\perp}$ forms 
	the set $\{a\ccc\mid a\in \F_q^*\}$ and all their supports are the set $\{s_1,s_2,\ldots,s_k\}$. 
	Therefore, each codeword of weight $k$ as well as its nonzero multiples in   $\C_{k}^{\perp}$ 
	with the support $\{s_1,s_2,\ldots,s_k\}$ correspond to  the set $\{x_1,x_2,\ldots,x_k\}$. 
	
	On one hand, by Equation (\ref{eq.subset problem}), the number of choices of $x_3,x_4,\ldots,x_k$ is independent of $x_1$ and $x_2$ 
	and it equals $\sum_{i=0}^{k-2}\sum_{j=0}^{k-2-i} (-1)^{i+j}N(i,j)$ and,
	on the other hand, by Equation (\ref{eq.N(i,j)}),  $N(i,j)$ is known for fixed $i$ and $j$. 
	Hence, $\sum_{i=0}^{k-2}\sum_{j=0}^{k-2-i} (-1)^{i+j}N(i,j)$ is known for a fixed $k$. 
	It then follows that the set of codewords of weight $k$ in $\C_k^{\perp}$ supports a $2$-$(q-1,k,\lambda_1)$ design, where 
	$$\lambda_1=\sum_{i=0}^{k-2}\sum_{j=0}^{k-2-i} (-1)^{i+j}N(i,j).$$
	%According to Lemma \ref{lm.t-design}, we further have $$A_k^{\perp}=\frac{\lambda_2(q-1)\binom{q-1}{t}}{\binom{k}{t}}.$$ 
	Furthermore, it turns out from Equation (\ref{eq.complement design}) that the set of minimum weight codewords of $\C_{k}$ supports 
	a $2$-$(q-1,q-k-1,\lambda_2)$ design, where 
	$$\lambda_2=\frac{\lambda_1\binom{q-3}{k}}{\binom{q-3}{k-2}},$$ 
and this completes the proof. 
\end{proof}

\begin{theorem}\label{th.conj333 weight distribution}
	The polynomial weight enumerators of $\C_{k}$ and   $\C_{k}^{\perp}$ are {given by} 
	$$A(z)=1+\sum_{i=q-k-1}^{n}A_iz^i~{\rm and}~A^{\perp}(z)=1+\sum_{i=k}^{n}A^{\perp}_iz^i,$$  {respectively,} 
	where $A_{q-k-1}=A^{\perp}_k=\frac{\lambda_1(q-1)^2(q-2)}{k(k-1)}$ and $\lambda_1=\sum_{i=0}^{k-2}\sum_{j=0}^{k-2-i} (-1)^{i+j}N(i,j).$ 
	Moreover, $A_i$ and $A^{\perp}_i$ are the same as those shown in Lemma \ref{lem.NMDS weight distribution}. 
\end{theorem}
\begin{proof}
	It follows from Equation (\ref{eq.parameters of t-design}), Lemma \ref{lem.complement design}, and  Theorem \ref{th.conj111} that 
	$$A_{q-k-1}=A^{\perp}_k=\frac{\lambda_1(q-1)\binom{q-1}{2}}{\binom{k}{2}}=\frac{\lambda_1(q-1)^2(q-2)}{k(k-1)}.$$ 
	Then the desired results follows straightforward from Lemma \ref{lem.NMDS weight distribution}. 
\end{proof}

Hereto, Conjecture~\ref{th.conj false} proposed by Heng $et~al.$ \cite{HW2023Conj} has been completely proved for $3\leq k\leq q-4$ 
and disproved for $k=q-3$ and $q-2$. 

In order to show our results in a more concrete way, we give the details for $k=4$ and $k=7$ in the form of {the following two} corollaries. Note that {the cases} $k=3,5$, and $6$ have been discussed   in Theorems 18, 27 and 35 of \cite{HW2023Conj}. 
It should be noticed that these two corollaries follow directly from Theorem~\ref{th.conj111} and Theorem~\ref{th.conj333 weight distribution} above.  
Moreover, we note that $\C_4$ has the same polynomial weight enumerator and support the same design as the one in  \cite[Theorem 23]{HW2023Conj}.

\begin{corollary}\label{coro.k=4}
	Let $q=2^h$ with $h\geq 3$. Then $\C_4$ is a $1$-MDS $[q-1,4,q-5]_q$ code with the polynomial weight enumerator 
	\begin{align*}
		\begin{split}
			A(z)= 1 &+\frac{(q-1)^2(q-2)(q-4)}{24}z^{q-5}+\frac{(q-1)^2(q-2)}{6}z^{q-4}+\frac{(q-1)^2(q-2)(q+4)}{4}z^{q-3} \\ 
			&  + \frac{(q-1)^2(2q^2+3q+28)}{6}z^{q-2}+\frac{(q-1)(9q^3+17q^2-18q+88)}{24}z^{q-1}. 
		\end{split}                
	\end{align*}
	Furthermore, the set of minimum weight codewords of $\C_4$ support a $2$-$(q-1,q-5,\frac{(q-4)(q-5)(q-6)}{24})$ design and 
	the set of minimum weight codewords of $\C_4^{\perp}$ support a $2$-$(q-1,4,\frac{q-4}{2})$ design.  
\end{corollary}

\begin{corollary}\label{coro.k=7}
	Let $q=2^h$ with $h\geq 4$. Then $\C_7$ is a $1$-MDS $[q-1,7,q-8]_q$ code with the polynomial weight enumerator 
	\begin{align*}
		\begin{split}
			A(z)= 1 & +\frac{(q-1)^2(q-2)(q-4)(q-6)(q^2-15q+71)}{5040}z^{q-8}\\ 
			& +\frac{7(q-1)^2(q-2)(q-4)(q-6)(q-8)}{720}z^{q-7} \\ 
			& + \frac{(q-1)^2(q-2)(q-4)(q^3-7q^2+49q-216)}{240}z^{q-6} \\ 
			& + \frac{(q-1)^2(q-2)(q-4)(2q^3+9q^2-26q+48)}{144}z^{q-5} \\ 
			& + \frac{(q-1)^2(q-2)(9q^4+11q^3-3q^2-26q-816)}{144}z^{q-4} \\ 
			& + \frac{(q-1)^2(q-2)(44q^4+155q^3+370q^2+280q+2496)}{240}z^{q-3} \\ 
			& + \frac{(q-1)^2(265q^5+663q^4+1075q^3+2430q^2-1880q+11712)}{720}z^{q-2} \\
			& + \frac{(q-1)(1854q^6+2779q^5+3423q^4+2380q^3+11676q^2-18704q+31872)}{5040}z^{q-1}. 
		\end{split}                
	\end{align*}
	Furthermore, the set of minimum weight codewords of $\C_7$ support a $2$-$(q-1,q-8,\frac{(q-4)(q-6)(q-8)(q-9)(q^2-15q+71)}{5040})$ design and 
	the set of minimum weight codewords of $\C_7^{\perp}$ support a $2$-$(q-1,7,\frac{(q-4)(q-6)(q^2-15q+71)}{120})$ design.  
\end{corollary}

\subsection{Infinite families of $1$-MDS optimal locally recoverable codes} 
Let $i\in \{1,2,\ldots,n\}$ and $R_i\subseteq \{1,2,\ldots,n\}\setminus \{i\}$ be a set of size $r$.  
Let $\ccc=(c_1,c_2,\ldots,c_n)\in \C$ and $\ccc_{R_i}$ be the projection of $\ccc$ at the positions in $R_i$. 
An $[n,k,d]_q$ linear code is called an {\textbf{$(n,k,d,q;r)$-locally recoverable code}} if for each $i$ there is a $R_i$ 
and a function $f_i(x_1,x_2,\ldots,x_r)$ on $\F_q^r$ such that $c_i=f_i(\ccc_{R_i})$.  
We call $R_i$ the {\textbf{recovering set}} of $c_i$ and, if $r$ is the minimum size of a recovering set or the {\textbf{minimum locality}} of $\C$,  then the $[n,k,d]_q$ linear code $\C$ is an $(n,k,d,q;r)$-locally recoverable code. 
 The following  well-known bounds apply for locally recoverable codes.

\begin{lemma}[\cite{CM2013}, {\textbf{Cadambe-Mazumdar bound}}]\label{lem.CM bound}
	Let $\C$ be an $(n,k,d,q;r)$-locally recoverable code and $k^{q}_{opt}(n,d)$ be the largest possible dimension of a $q$-ary linear code. 
	Suppose that $\mathbb{Z}^{+}$ is the set of all positive integers. Then 
	\begin{align}
		k\leq \min_{t\in \mathbb{Z}^{+}}\{rt+k^q_{opt}(n-t(r+1),d)\}.  
	\end{align}
\end{lemma}

\begin{lemma}[\cite{GHSY2012}, {\textbf{Singleton-like bound}}]\label{lem.Singleton-like bound}
	Let $\C$ be an $(n,k,d,q;r)$-locally recoverable code. Then 
	\begin{align}
		d\leq n-k-\left\lceil \frac{k}{r} \right\rceil+2. 
	\end{align}
\end{lemma}

An $(n,k,d,q;r)$-locally recoverable code is called {\textbf{$k$-optimal}} (resp. {\textbf{$d$-optimal}}) 
if it achieves the Cadambe-Mazumdar bound (resp. the Singleton-like bound). 
Let $\BBB_i(\C)=\{\Supp(\ccc)\mid \ccc\in \C~{\rm and}~\wt(\ccc)=i\}$,
Tan $et~al.$ \cite{TFDTZ2023} proposed the following result, which is important for us to determine the minimum locality of a $1$-MDS code.

\begin{lemma}[Corollary 3 in \cite{TFDTZ2023}]\label{lem.locality}
	Let $\C$ be a nontrivial linear code with length $n$. If $d(\C^{\perp})=d^{\perp}$ and $(\PP(\C^{\perp}),\BBB_{d^{\perp}}(\C^{\perp}))$ is 
	a $1$-$(n,d^{\perp},\lambda^{\perp})$ design with $\lambda^{\perp}\geq 1$, then $\C$ has minimum locality $d^{\perp}-1$. 
\end{lemma}

\begin{theorem}\label{th.optimal LRC}
	Let $q=2^h$ with $h\geq 3$ and $3\leq k\leq q-4$. Then the following statements hold. 
	\begin{enumerate}
		\item   The linear code $\C_k$ is a $k$-optimal and $d$-optimal $(q-1,k,q-k-1,q,k-1)$-locally recoverable code. 
		
		\item  The linear code $\C_k^{\perp}$ is a $k$-optimal and $d$-optimal $(q-1,q-k-1,k,q,q-k-2)$-locally recoverable code. 
	\end{enumerate}
\end{theorem}
\begin{proof}
	We prove the result 1) and the proof of the result 2) is similar. 
	It follows from Theorem \ref{th.conj111} and Lemma \ref{lem.locality} that $\C_k$ has minimum locality $k-1$. 
	Note that 
	$k^q_{opt}(q-1-t(k-1+1),q-k-1)=k^q_{opt}(q-tk-1,q-k-1).$ 
	From the Singleton bound,  we have $k^q_{opt}(q-tk-1,q-k-1)=k(1-t)+1$ and $t\leq 1$. 
	Hence, $$\min_{t\in \mathbb{Z}^{+}}\{rt+k^q_{opt}(n-t(r+1),d)\}=(k-t+1)|_{t=1}=k.$$
	%By Lemma \ref{lem.CM bound}, $\C_k$ is $k$-optimal.  
	Note also that 
	$$q-k-1-\left\lceil \frac{k}{k-1} \right\rceil+2=q-k-1.$$ 
	Therefore, the desired results follow from Lemmas \ref{lem.CM bound} and \ref{lem.Singleton-like bound}.  
\end{proof}

%\subsection{Constructions of $m$-MDS codes}\label{sec4}
\section{New $\ell$-MDS codes from old ones}\label{sec4}

%$$\textbf{Can new $m'$-MDS codes be constructed from known $m$-MDS codes?}$$ 
In this section, we construct new $\ell$-MDS codes from old ones. The  {objective for that is} two-folded,  
 first  to study the $\ell$-MDS properties of extended and expurgated codes of a known binary $\ell_1$-MDS code;
second  to derive the $\ell$-MDS properties of linear codes arising from the $(\uuu,\uuu+\vvv)$ construction of two known $\ell_1$-MDS and $\ell_2$-MDS codes. 
%As a result, many $\ell$-MDS codes with larger $\ell$ can be obtained. Therefore, these new $\ell$-MDS codes can be applied for index coding problems and informed source coding problems even if the original ones can not. 
It has been emphasized that according to Lemma \ref{lem.m-MDS sufficient condition} the construction of $\ell$-MDS codes generally becomes more difficult as $\ell$ increases.

\subsection{New $\ell$-MDS codes from extended and expurgated codes}\label{sec4.1} 

In this subsection, we focus on the binary case, i.e., linear codes over $\mathbb F_2$. 
First, we study the $\ell$-MDS property of the extended code of an $\ell_1$-MDS code. 
%To this end, we need the following definitions and results. 

\begin{definition}[\cite{MS1977}]\label{def.extended code}
	Let $\C$ be an $[n,k,d]_2$ linear code. The {\textbf{extended code}} of $\C$ is the linear code $\EE(\C)$ 
	defined as
	\begin{align}
		\EE(\C)=\left\{(c_1,c_2,\ldots,c_{n+1})\in \F_q^{n+1}\mid (c_1,c_2,\ldots,c_n)\in \C~{\rm with}~\sum_{i=1}^{n+1}c_i=0\right\}, 
	\end{align}
	 and it is an $[n+1,k,d(\EE(\C))]_2$ code, where $d(\EE(\C))=d$ if $d$ is even and $d(\EE(\C))=d+1$ if $d$ is odd. 
\end{definition}

\begin{definition}[\cite{MS1977}]\label{def.augmented code}
	Let $\C$ be an $[n,k,d]_2$ linear code and the all-one vector $\mathbf{1}\notin \C$.  
	The  {\textbf{augmented code}} of $\C$ is the linear code $\AAA(\C)$
	defined as
	\begin{align}
		\AAA(\C)=\C\cup \{\mathbf{1}+\C\}, 
	\end{align}
	and it is  an $[n,k+1,\min\{d,n-d'\}]_2$, where $d'$ is the largest weight of all the codewords in $\C$. 
\end{definition}

It is clear that, if the all-one vector $\mathbf{1}\in \C^{\perp}$, 
then binary linear code $\EE(\C)$ is trivial. 
Hence, we will assume that $\mathbf{1}\notin \C^{\perp}$  in the sequel. 
The following result provides the minimum distance of the dual code of a non-trivial extended code. 
\begin{lemma}[Theorem 12 in \cite{SDC2023}]\label{lem.parameters of extended code}
	Let $\C$ be an $[n,k,d]_2$ linear code with dual distance $d^{\perp}$.
	Suppose that $\mathbf{1}\notin \C^{\perp}$. 
	Then 
	\begin{align}
		d(\EE(\C)^{\perp})= \left\{\begin{array}{lr}
			d(\AAA(\C^{\perp}))+1, & {\rm if}~d(\AAA(\C^{\perp}))<d^{\perp}, \\
			d^{\perp}, & {\rm if}~d(\AAA(\C^{\perp}))\geq d^{\perp}. 
		\end{array}   \right.
	\end{align}
\end{lemma}

\begin{theorem}\label{th.m-MDS code from extended codes}
	Let $\C$ be an $\ell_1$-MDS $[n,k]_2$ code.  
	Let $d'^{\perp}$ be the largest weight of codewords in $\C^{\perp}$.  
	Suppose that $\mathbf{1}\notin \C^{\perp}$. Then the following statements hold. 
	\begin{enumerate}
		\item  If $n-k-\ell_1$ is odd, then $\EE(\C)$ is an $[n+1,k,n-k-\ell_1+1]_q$ code 
		and $\EE(\C)$ is an $\ell$-MDS code if and only if $d'^{\perp}=n-k+\ell_1+1$, if and only if $\ell=\ell_1+1$. 
		
		\item  If $n-k-\ell_1$ is even, then $\EE(\C)$ is an $[n+1,k,n-k-\ell_1+2]_q$ code 
		and $\EE(\C)$ is an $\ell$-MDS code if and only if $d'^{\perp}\leq n-k+\ell_1$, if and only if $\ell=\ell_1$.  
	\end{enumerate}
\end{theorem}
\begin{proof}
	Since $\C$ is an $\ell_1$-MDS $[n,k]_2$ code, then $\C$ and $\C^{\perp}$ 
	have respective parameters $[n,k,n-k-\ell_1+1]_2$ and $[n,n-k,k-\ell_1+1]_2$.  
	By Definition \ref{def.augmented code}, $d(\AAA(\C^{\perp}))=\min\{k-\ell_1+1,n-d'^{\perp}\}$.  
	Note that $d(\AAA(\C^{\perp}))<k-\ell_1+1$ holds if and only if $n-d'^{\perp}<k-\ell_1+1$, i.e., $d'^{\perp}>n-k+\ell_1-1$ holds. 
	Hence, taking into account Lemma~\ref{lem.parameters of extended code}, we have that 
	\begin{align}\label{eq.dual extended code distance}
		d(\EE(\C)^{\perp})= \left\{\begin{array}{lr}
			n-d'^{\perp}+1, & {\rm if}~d'^{\perp}>n-k+\ell_1-1, \\
			k-\ell_1+1, & {\rm if}~d'^{\perp}\leq n-k+\ell_1-1. 
		\end{array}   \right.
	\end{align}

	\begin{enumerate}
		\item 
	
	 Since $n-k-\ell_1$ is odd, then $n-k-\ell_1+1$ is even and it follows from Definition \ref{def.extended code}  
	that $\EE(\C)$ has parameters $[n+1,k,n-k-\ell_1+1]_2$. Then we have $\SSS(\EE(\C))=\ell_1+1$ and the following two cases. 
	\begin{description}
		\item{\bf{Case 1}.-} If $d'^{\perp}\leq n-k+\ell_1-1$, it turns out from Equation (\ref{eq.dual extended code distance})  
		that $\EE(\C)^{\perp}$ has parameters $[n+1,n-k+1,k-\ell_1+1]_2$, which implies that $\SSS(\EE(\C)^{\perp})=\ell_1$. 
		Note that $\SSS(\EE(\C))\neq \SSS(\EE(\C)^{\perp})$ for each positive integer $\ell_1$. 
		Hence, $\EE(\C)$ is not an $\ell$-MDS code.   
		
		\item{\bf{Case 2}.-} If $d'^{\perp}> n-k+\ell_1-1$, it turns out from Equation (\ref{eq.dual extended code distance}) again 
		that $\EE(\C)^{\perp}$ has parameters $[n+1,n-k+1,n-d'^{\perp}+1]_2$. 
		It implies that $\SSS(\EE(\C)^{\perp})=k-n+d'^{\perp}$. 
		Note that $\SSS(\EE(\C))=\SSS(\EE(\C)^{\perp})$ if and only if $d'^{\perp}=n-k+\ell_1+1$, 
		if and only if $\EE(\C)$ is an $(\ell_1+1)$-MDS code.  
	\end{description}

	\item Since $n-k-\ell_1$ is even, then $n-k-\ell_1+1$ is odd and it follows from Definition \ref{def.extended code}  
	that $\EE(\C)$ has parameters $[n+1,k,n-k-\ell_1+2]_2$. 
	Then we have $\SSS(\EE(\C))=\ell_1$ and the following two cases. 
	
	\begin{description}
		\item{\bf{Case 1}.-}  If $d'^{\perp}\leq n-k+\ell_1-1$,  
		from the discussion of 1) above, we have $\SSS(\EE(\C)^{\perp})=\ell_1$. 
		Hence, $\EE(\C)$ is an $\ell$-MDS code if and only if $\ell=\ell_1$ in this case.   
		
		\item{\bf{Case 2}.-} If $d'^{\perp}> n-k+\ell_1-1$, from the discussion of 1) above again, 
		we have $\SSS(\EE(\C)^{\perp})=k-n+d'^{\perp}$. 
		Note that $\SSS(\EE(\C))=\SSS(\EE(\C)^{\perp})$ if and only if $d'^{\perp}=n-k+\ell_1$, 
		if and only if $\EE(\C)$ is an $\ell_1$-MDS code.  
	\end{description}
	
\end{enumerate}
\end{proof}

Next, we study the $\ell$-MDS property of the expurgated code of an $\ell_1$-MDS code. 
%The expurgated code can be defined as follows. 

\begin{definition}[\cite{MS1977}]\label{def.expurgated code}
	Let $\C$ be an $[n,k,d]_2$ linear code with at least one odd weight codeword.    
	The {\textbf{expurgated code}} of $\C$ is linear code $\widetilde{\C}$ 
	given by 
	\begin{align}
		\widetilde{\C}=\{\ccc\mid \ccc\in \C~{\rm and}~\wt(\ccc)\equiv 0~({\rm mod}~2)\}, 
	\end{align} and it is an $[n,k-1,d(\widetilde{\C})]_2$
	where $d(\widetilde{\C})\geq d$ and the equality holds if $d$ is even.  
\end{definition}

%Then we have the following theorem. 

\begin{theorem}\label{th.m-MDS code from expurgated codes}
	Let $\C$ be an $\ell$-MDS $[n,k]_2$ code, where $n-k-\ell_1$ is odd.   
	Let $d'^{\perp}$ be the largest weight of all the codewords in $\C^{\perp}$.  
	Suppose that $\mathbf{1}\notin \C^{\perp}$. 
	Then $\widetilde{\C}$ is an $[n,k-1,n-k-\ell_1+1]_2$ code and 
	$\widetilde{\C}$ is an $\ell$-MDS code if and only if $d'^{\perp}=n-k+\ell_1+1$, if and only if $\ell=\ell_1+1$.  
\end{theorem}
\begin{proof}
	Since $\C$ is an $\ell_1$-MDS $[n,k]_2$ code, then $\C$ and $\C^{\perp}$ have parameters $[n,k,n-k-\ell_1+1]_2$ and $[n,n-k,k-\ell_1+1]_2$, respectively. 
	On one hand, since $n-k-\ell_1$ is odd, it follows from Definition \ref{def.expurgated code} that $\widetilde{\C}$ is an $[n,k-1,n-k-\ell_1+1]_2$ code, 
	which further implies that $\SSS(\widetilde{\C})=\ell_1+1$. 
	On the other hand, it is not difficult to see that $(\widetilde{\C})^{\perp}=\C^{\perp}\cup \{\mathbf{1}+\C^{\perp}\}$ (see also\cite{HYUS2016}). 
	Combining Definition \ref{def.augmented code}, if $\mathbf{1}\notin \C^{\perp}$, we immediately have $\AAA(\C^{\perp})=\C^{\perp}\cup \{\mathbf{1}+\C^{\perp}\}$, 
	which implies that $(\widetilde{\C})^{\perp}=\AAA(\C^{\perp})$ is an 
	$[n,n-k+1,\min\{k-\ell_1+1,n-d'^{\perp}\}]_2$ code. 
	We have the following two cases. 
	\begin{description}
		\item[]\textbf{Case 1}.- If $d'^{\perp}\leq n-k+\ell_1-1$, then 
	$(\widetilde{\C})^{\perp}$ is an $[n,n-k+1,k-\ell_1+1]_2$ code and hence, $\SSS((\widetilde{\C})^{\perp})=\ell_1-1$. 
	Note that $\SSS(\widetilde{\C})\neq \SSS((\widetilde{\C})^{\perp})$ for each positive integer $\ell_1$.  
	Hence, $\widetilde{\C}$ is not an $\ell$-MDS code. 
	
	\item[]\textbf{Case 2}.- If $d'^{\perp}> n-k+\ell_1-1$, then 
	$(\widetilde{\C})^{\perp}$ is an $[n,n-k+1,n-d'^{\perp}]_2$ code and hence, $\SSS((\widetilde{\C})^{\perp})=k+d'^{\perp}-n$. 
	Note that $\SSS(\widetilde{\C})=\SSS((\widetilde{\C})^{\perp})$ if and only if $d'^{\perp}=n-k+\ell_1+1$, 
	if and only if $\widetilde{\C}$ is an $(\ell_1+1)$-MDS code.  
	\end{description}	
This completes the proof. 
\end{proof}

\begin{example}\label{exam.extended Hamming codes}
	With \cite[Theorem 3.8]{LL2014-m-MDS-Liao}, one knows the binary Hamming code $\Ham(r,2)$ 
	with parameters $[2^r-1,2^r-r-1,3]_2$ is an $\ell$-MDS code if and only if $r=2$ or $3$. More precisely, 
	$\Ham(2,2)$ is an MDS $[3,1,3]_2$ code and  $\Ham(3,2)$ is a $1$-MDS $[7,4,3]_2$ code. 
	Note that $\Ham(2,2)^{\perp}$ and $\Ham(3,2)^{\perp}$ are $[3,2,2]_2$ and $[7,3,4]_2$ 
	simplex codes, respectively. Then the largest weights of codewords in $\Ham(2,2)^{\perp}$ 
	and $\Ham(3,2)^{\perp}$ are respectively $2$ and $4$, which also implies that 
	$\mathbf{1}\notin  \Ham(2,2)^{\perp}$ and $\mathbf{1}\notin  \Ham(3,2)^{\perp}$. 
	It then follows from Theorem \ref{th.m-MDS code from extended codes} that $\EE(\Ham(2,2))$ is an 
	MDS $[4,1,4]_2$ code and $\EE(\Ham(3,2))$ is a $1$-MDS $[8,4,4]_2$ code. 
	Furthermore, one has that $\EE(\Ham(2,2))^{\perp}$ is an 
	MDS $[4,3,2]_2$ code and $\EE(\Ham(3,2))^{\perp}$ is again a $1$-MDS $[8,4,4]_2$ code. 
	Note also that all these $\ell$-MDS codes are optimal \cite{codetable}. 
\end{example}

\begin{example}\label{exam.m-MDS codes from extended codes}
	We list in Table \ref{tab:1} some $\ell$-MDS codes obtained from 
	extended codes and expurgated codes of initial codes as well as their dual codes named 
	respectively dual extended codes and dual expurgated codes, where the initial codes are 
	selected from the current $\textsc{Magma}$ $\texttt{BKLC}$ database \cite{magma,codetable}.  
	We denote resulting optimal codes or best-known linear codes in bold and 
	mark derived almost optimal codes or almost best-known linear codes with a $^\star$. 
	In addition, we have also verified by $\textsc{Magma}$ \cite{magma} that these derived $\ell$-MDS codes are not 
	monomially equivalent to linear codes in the current $\textsc{Magma}$ $\texttt{BKLC}$ database \cite{magma,codetable}, 
	that is, they are new in the sense of monomial equivalence.  
	
	\begin{table}
		%\footnotesize
		\centering
		\caption{Some $\ell$-MDS codes from Theorems \ref{th.m-MDS code from extended codes} and \ref{th.m-MDS code from expurgated codes}}
		\label{tab:1}       % Give a unique label
		\begin{tabular}{c|c|c|c|c}
			\hline
			Initial code & Extended code & Dual extended code & $\ell$-MDS property  & Reference \\ \hline\hline
			
			$[43,21,10]_2$ & $[44,21,10]_2^\star$ & $[44,23,8]_2^\star$ & $14$-MDS &  Theorem \ref{th.m-MDS code from extended codes} 1) \\ 
			
			$[44,22,10]_2$ & $[45,22,10]_2^\star$ & $[45,23,9]_2^\star$ & $14$-MDS &  Theorem \ref{th.m-MDS code from extended codes} 1) \\ 
			
			%$[45,23,10]_2$ & $[46,23,10]_2^\star$ & $[46,23,10]_2^\star$ & $14$-MDS &  Theorem \ref{th.m-MDS code from extended codes} 1) \\ 
			
			$[46,24,10]_2$ & $[47,24,10]_2^\star$ & $[47,23,11]_2^\star$ & $14$-MDS &  Theorem \ref{th.m-MDS code from extended codes} 1) \\ \hline
			
			Initial code & Extended code & Dual extended code & $\ell$-MDS property  & Reference \\ \hline\hline
			
			$[13,5,5]_2$ & $\mathbf{[14,5,6]_2}$ & $[14,9,2]_2$ & $4$-MDS &  Theorem \ref{th.m-MDS code from extended codes} 2) \\
			
			$[37,17,9]_2$ & $\mathbf{[38,17,10]_2}$ & $[38,21,6]_2$ & $9$-MDS &  Theorem \ref{th.m-MDS code from extended codes} 2) \\
			
			%$[41,18,11]_2$ & $\mathbf{[42,18,12]_2}$ & $[42,24,6]_2$ & $13$-MDS &  Theorem \ref{th.m-MDS code from extended codes} 2) \\ 
			
			$[42,19,11]_2$ & $\mathbf{[43,19,12]_2}$ & $[43,24,7]_2^\star$ & $13$-MDS &  Theorem \ref{th.m-MDS code from extended codes} 2) \\ 
			
			%$[43,20,11]_2$ & $\mathbf{[44,20,12]_2}$ & $\mathbf{[44,24,8]_2}$ & $13$-MDS &  Theorem \ref{th.m-MDS code from extended codes} 2) \\ 

			$[59,26,13]_2$ & $\mathbf{[60,26,14]_2}$ & $[60,34,6]_2$ & $21$-MDS &  Theorem \ref{th.m-MDS code from extended codes} 2) \\ \hline
			
			Initial code & Expurgated code & Dual expurgated code & $\ell$-MDS property  & Reference \\ \hline \hline
			
			$[43,21,10]_2$ & $[43,20,10]_2^\star$ & $[43,23,7]_2^\star$ & $14$-MDS &  Theorem \ref{th.m-MDS code from expurgated codes} \\ 
			
			%$[44,22,10]_2$ & $[44,21,10]_2^\star$ & $[44,23,8]_2^\star$ & $14$-MDS &  Theorem \ref{th.m-MDS code from expurgated codes} \\ 
			
			$[45,23,10]_2$ & $[45,22,10]_2^\star$ & $\mathbf{[45,23,9]_2}$ & $14$-MDS &  Theorem \ref{th.m-MDS code from expurgated codes} \\ 
			
			$[46,24,10]_2$ & $[46,23,10]_2^\star$ & $[46,23,10]_2^\star$ & $14$-MDS &  Theorem \ref{th.m-MDS code from expurgated codes} \\ \hline
		\end{tabular}
	\end{table}
	
\end{example}

\subsection{New $\ell$-MDS codes from the $(\uuu,\uuu+\vvv)$ construction}\label{sec4.2}

\begin{definition}\label{def.(u,u+v)}
	Let $\C_i$ be an $[n,k_i]_q$ linear code for $i=1, 2$. 
	%Let $G_i$ and $H_i$ be respectively a generator matrix and a parity check matrix of $\C_i$ for $i=1, 2$. 
	The {\textbf{$(\uuu,\uuu+\vvv)$ construction}} of $\C_1$ and $\C_2$ is the linear code $\PP(\C_1,\C_2)$ defined as  
	\begin{align}
		\PP(\C_1,\C_2)=\{(\mathbf{u},\mathbf{u+v})\mid \mathbf{u}\in \C_1,\ \mathbf{v}\in \C_2\},    
	\end{align}
	%where $d(\PP(\C_1,\C_2))=\min\{2d_1,d_2\}.$
	%whose generator matrix and parity check matrix are given by  
	%\begin{align}\label{eq.(u,u+v) matrix}
	%    G=\left(\begin{array}{cc}
		%        G_1 & G_1 \\
		%        O_{k_2\times n} & G_2 
		%    \end{array}\right)\ {\rm and}\  H=\left(\begin{array}{cc}
		%        H_1 & O_{(n-k_1)\times n} \\
		%        -H_2 & H_2 
		%    \end{array}\right).
	%\end{align}
and it is a $[2n,k_1+k_2,\min\{2d_1,d_2\}]_q$ code.
\end{definition}

\begin{definition}\label{def.FSD codes}
	Let $\C$ be a $q$-ary linear code. If $\C$ and $\C^{\perp}$ have the same weight distribution, 
	$\C$ is called a {\textbf{formally self-dual (FSD)}} code. 
\end{definition}

With above definition, it is clear that FSD codes are $\ell$-MDS codes and contain self-dual codes ($\C=\C^{\perp}$) as a special subclass. 
For more details on FSD codes, one can refer to \cite{KP1994,LSL2023,LSW2023,LZM2023} and the references therein.

\begin{lemma}\label{prop.parameters of (u,u+v)}
	Let $\C_i$ be an $[n,k_i,d_i]_q$ linear code with dual distance $d_i^{\perp}$ for $i=1, 2$. 
	Then $(\PP(\C_1,\C_2))^{\perp}$ is monomially equivalent to $\PP(\C_2^{\perp},\C_1^{\perp})$. 
	Moreover, $(\PP(\C_1,\C_2))^{\perp}$ has parameters $[2n,2n-k_1-k_2,\min\{d_1^{\perp}, 2d_2^{\perp}\}]_q$. 
\end{lemma}
\begin{proof}
	For $q=2$,  it has been determined in \cite{MS1977} that 
	$(\PP(\C_1,\C_2))^{\perp}=\{(\aaa+\bbb,\bbb)\mid \aaa\in \C_1^{\perp},\bbb\in \C_2^{\perp}\}$. 
	%For general $q$, we similarly have $(\PP(\C_1,\C_2))^{\perp}=\{(\aaa+\bbb,-\bbb)\mid \aaa\in \C_1^{\perp},\bbb\in \C_2^{\perp}\},$  
	%which is monomially equivalent to $\PP(\C_2^{\perp},\C_1^{\perp})=\{(\bbb,\aaa+\bbb)\mid \aaa\in \C_1^{\perp},\bbb\in \C_2^{\perp}\}$.  
	For general $q$, it can be verified that 
	\begin{align*}
		(\PP(\C_1,\C_2))^{\perp} & = \{(\aaa+\bbb,-\bbb)\mid \aaa\in \C_1^{\perp},\bbb\in \C_2^{\perp}\} \\
		& \simeq \{(\bbb,\aaa+\bbb)\mid \aaa\in \C_1^{\perp},\bbb\in \C_2^{\perp}\} \\
		& = \PP(\C_2^{\perp},\C_1^{\perp}), 
	\end{align*}
	which ``$\simeq$'' denotes the monomial equivalence. 
	Moreover, the parameters of $(\PP(\C_1,\C_2))^{\perp}$ follows straightforward from the monomial equivalence and Definition \ref{def.(u,u+v)}, which  completes the proof. 
\end{proof}

%In the following, we consider a special case where $\C_1=\C_2$ are $m$-MDS codes. 

\begin{theorem}\label{th.m-MDS codes from C and C dual}
	%Let $\C_1$ and $\C_2$ be $[n,\frac{n}{2}]_q$ $m$-MDS codes with even $n$. 
	%Then $\PP(\C_1,\C_2)$ is an $(\frac{n}{2}+m)$-MDS $[2n,n,\frac{n}{2}-m+1]_q$ linear code. 
	Let $\C$ be an $[n,k,d]_q$ linear code with dual distance $d^{\perp}$. 
	Then the following statements hold. 
	\begin{enumerate}
		\item  $\PP(\C,\C^{\perp})$ is a 
		$[2n,n,\min\{2d,d^{\perp}\}]_q$ FSD code. 
		
		\item If $\C$ is an $\ell$-MDS code with $\ell\geq 2n-3k+1$, 
		then $\PP(\C,\C^{\perp})$ is a $(2k+2\ell-n-1)$-MDS $[2n,n,2n-2k+2-2\ell]_q$ code. 
		
		\item  If $\C$ is an $\ell$-MDS code with $\ell< 2n-3k+1$, 
		then $\PP(\C,\C^{\perp})$ is an $(n-k+\ell)$-MDS $[2n,n,k+1-\ell]_q$ code. 
	\end{enumerate}
\end{theorem}
\begin{proof} $\quad$
	%From Theorem \ref{th.m-MDS codes from (u,u+v) construction} 4), the desired result is straightforward. 
	\begin{enumerate}
		\item[1)] From Definition \ref{def.(u,u+v)}, 
	$\PP(\C,\C^{\perp})$ has the desired parameters. % $[2n,n,\min\{2d,d^{\perp}\}]_q$. 
	By Lemma \ref{prop.parameters of (u,u+v)}, 
	$(\PP(\C,\C^{\perp}))^{\perp}$ is monomially equivalent to 
	$\PP((\C^{\perp})^{\perp},\C^{\perp})=\PP(\C,\C^{\perp})$. 
	Since monomially equivalent linear codes have the same weight distribution, $\PP(\C,\C^{\perp})$ is FSD.  
	%This completes the proof of the result 1). 
	
	\item[2-3)] Since $\C$ is an $\ell$-MDS code, so does $\C^{\perp}$. 
	Then with 1) above, $\PP(\C,\C^{\perp})$ is a $[2n,n,\min\{2n-2k-2\ell+2,k-\ell+1\}]_q$ FSD code. 
	More precisely, $\PP(\C,\C^{\perp})$ is a $[2n,n,2n-2k-2\ell+2]_q$ FSD code if $\ell\geq 2n-3k+1$ 
	and a $[2n,n,k-\ell+1]_q$ FSD code if $\ell< 2n-3k+1$.  
	Hence, the desired results 2) and 3) clearly holds. 
\end{enumerate}
\end{proof}

\begin{corollary}\label{coro.m-MDS codes from (u,u+v) MDS codes}
	Let $q=2^h$ with $h\geq 3$. Then the following statements hold. 
	\begin{enumerate}
		\item  There exists a $(q-4)$-MDS $[2q-2,q-1,4]_q$ FSD code.  
		\item  There exists a $(q-6)$-MDS $[2q-2,q-1,6]_q$ FSD code.  
	\end{enumerate} 
\end{corollary}
\begin{proof}
	From Theorem \ref{th.conj false}, we take $\C_1$ as an MDS $[q-1,q-2,2]_q$ code and $\C_2$ as an MDS $[q-1,q-3,3]_q$ code.  
	Consider the linear codes $\PP(\C_1,\C_1^{\perp})$ and $\PP(\C_2,\C_2^{\perp})$. 
	Since $q=2^h\geq 8$, 1) and 2) follow from Theorem \ref{th.m-MDS codes from C and C dual}. 
\end{proof}

\begin{corollary}\label{coro.l-MDS codes from conj111}
	Let $q=2^h$ with $h\geq 3$ and $3\leq k\leq q-4$. 
	Then the following statements hold. 
	\begin{enumerate}
		\item  If $3\leq k\leq \lfloor \frac{q-1}{3} \rfloor$, then there exists a $(q-2k)$-MDS $[2q-2,q-1,2k]_q$ FSD code. 
		\item   If $\lceil \frac{q-1}{3} \rceil \leq k\leq q-4$, then there exists a $(k+1)$-MDS $[2q-2,q-1,q-k-1]_q$ FSD code. 
	\end{enumerate}
\end{corollary}
\begin{proof}
	From Theorem \ref{th.conj111}, we take $\C$ as a $1$-MDS $[q-1,k,q-k-1]_q$ code. 
	Consider the linear code $\PP(\C^{\perp},\C)$, and the corollary follows from Theorem \ref{th.m-MDS codes from C and C dual}. 
\end{proof}

\begin{theorem}\label{th.m-MDS codes from (u,u+v) construction}
	Let $\C_1$ be an $\ell_1$-MDS $[n,k_1]_q$ code and $\C_2$ be an $\ell_2$-MDS $[n,k_2]_q$ code. 
	Then the following statements hold. 
	\begin{enumerate}
		\item   If $\lceil \frac{2k_2-k_1+\ell_1+1}{2} \rceil\leq \ell_2\leq 2k_1-k_2+2\ell_1-n-1$, 
		then $\PP(\C_1,\C_2)$ is a $[2n,k_1+k_2,2n-2k_1-2\ell_1+2]_q$ linear code and $\PP(\C_1,\C_2)$ is an $\ell$-MDS code 
		if and only if $\ell_1=\ell_2$, if and only if $\ell=k_1-k_2+2\ell_1-1$. 
		
		\item   If $\ell_2\leq \min\{\lfloor \frac{2k_2-k_1+\ell_1+1}{2} \rfloor, 2k_1-k_2+2\ell_1-n-1\}$, 
		then $\PP(\C_1,\C_2)$ is a $[2n,k_1+k_2,2n-2k_1-2\ell_1+2]_q$ linear code and $\PP(\C_1,\C_2)$ is an $\ell$-MDS code 
		if and only if $\ell_1=2k_2-k_1+1$, if and only if $\ell=3k_2-k_1+1$. 
		
		\item  If $\ell_2\geq \max\{\lceil \frac{2k_2-k_1+\ell_1+1}{2} \rceil, 2k_1-k_2+2\ell_1-n-1\}$, 
		then $\PP(\C_1,\C_2)$ is a $[2n,k_1+k_2,n-k_2-\ell_2+1]_q$ linear code and $\PP(\C_1,\C_2)$ is an $\ell$-MDS code 
		if and only if $\ell_2=n-2k_1+k_2+1$, if and only if $\ell=2n-3k_1+k_2+1$. 
		
		\item   If $2k_1-k_2+2\ell_1-n-1\leq \ell_2\leq \lfloor \frac{2k_2-k_1+\ell_1+1}{2} \rfloor$,  
		then $\PP(\C_1,\C_2)$ is a $[2n,k_1+k_2,n-k_2-\ell_2+1]_q$ linear code and $\PP(\C_1,\C_2)$ is an $\ell$-MDS code 
		if and only if $\ell_2=k_1+k_2+\ell_1-n$, if and only if $\ell=k_2+\ell_1$. 
	\end{enumerate}
\end{theorem}
\begin{proof}
	Since $\C_1$ is $\ell_1$-MDS and $\C_2$ is $\ell_2$-MDS, it follows from Definition \ref{def.(u,u+v)} and Lemma \ref{prop.parameters of (u,u+v)} that 
	$\PP(\C_1,\C_2)$ and $(\PP(\C_1,\C_2))^{\perp}$ have  parameters $[2n,k_1+k_2,\min\{2n-2k_1-2\ell_1+2,n-k_2-\ell_2+1\}]_q$ 
	and $[2n,2n-k_1-k_2,\min\{2k_2-2\ell_2+2,k_1-\ell_1+1\}]_q$ respectively. 
	\begin{enumerate}
		\item 
	 If $2n-2k_1-2\ell_1+2\leq n-k_2-\ell_2+1$ and $2k_2-2\ell_2+2\leq k_1-\ell_1+1$, i.e.,  
	$\lceil \frac{2k_2-k_1+\ell_1+1}{2} \rceil\leq \ell_2\leq 2k_1-k_2+2\ell_1-n-1$, 
	then $\PP(\C_1,\C_2)$ has parameters  $[2n,k_1+k_2,2n-2k_1-2\ell_1+2]_q$ and 
	$(\PP(\C_1,\C_2))^{\perp}$ has parameters $[2n,2n-k_1-k_2,2k_2-2\ell_2+2]_q$. 
	It implies that $\SSS(\PP(\C_1,\C_2))=k_1-k_2+2\ell_1-1$ and $\SSS((\PP(\C_1,\C_2))^{\perp})=k_1-k_2+2\ell_2-1$. 
	Note that $\SSS(\PP(\C_1,\C_2))=\SSS(\PP(\C_1,\C_2)^{\perp})$ if and only if $\ell_1=\ell_2$, 
	if and only if $\PP(\C_1,\C_2)$ is a $(k_1-k_2+2\ell_1-1)$-MDS code.

	\item If $2n-2k_1-2\ell_1+2\leq n-k_2-\ell_2+1$ and $2k_2-2\ell_2+2\geq k_1-\ell_1+1$, i.e.,  
	$\ell_2\leq \min\{\lfloor \frac{2k_2-k_1+\ell_1+1}{2} \rfloor, 2k_1-k_2+2\ell_1-n-1\}$, 
	then $\PP(\C_1,\C_2)$ has parameters  $[2n,k_1+k_2,2n-2k_1-2\ell_1+2]_q$ and 
	$(\PP(\C_1,\C_2))^{\perp}$ has parameters $[2n,2n-k_1-k_2,k_1-\ell_1+1]_q$. 
	It implies that $\SSS(\PP(\C_1,\C_2))=k_1-k_2+2\ell_1-1$ and $\SSS((\PP(\C_1,\C_2))^{\perp})=k_2+\ell_1$. 
	Note that $\SSS(\PP(\C_1,\C_2))=\SSS(\PP(\C_1,\C_2)^{\perp})$ if and only if $\ell_1=2k_2-k_1+1$, 
	if and only if $\PP(\C_1,\C_2)$ is a $(3k_2-k_1+1)$-MDS code. 
	
	\item  If $2n-2k_1-2\ell_1+2\geq n-k_2-\ell_2+1$ and $2k_2-2\ell_2+2\leq k_1-\ell_1+1$, i.e.,  
	$\ell_2\geq \max\{\lceil \frac{2k_2-k_1+\ell_1+1}{2} \rceil, 2k_1-k_2+2\ell_1-n-1\}$, 
	 and by a similar argument as 1) and 2) above, we get that  the  result 3) holds.

	\item If $2n-2k_1-2\ell_1+2\geq n-k_2-\ell_2+1$ and $2k_2-2\ell_2+2\geq k_1-\ell_1+1$, i.e.,  
	$2k_1-k_2+2\ell_1-n-1\leq \ell_2\leq \lfloor \frac{2k_2-k_1+\ell_1+1}{2} \rfloor$,  and  4) follows immediately by similar discussions as above. 
\end{enumerate}
\end{proof}

\begin{corollary}\label{coro.l-MDS codes from conj222}
	Let $q=2^h$ with $h\geq 3$ and $3\leq k_1, k_2\leq q-4$. 
	If $\max\{2k_1+k_2, k_1+2k_2\} \leq q-1$, 
	then there exists a $(q-k_1-k_2)$-MDS $[2q-2,q+k_1-k_2-1,2k_2]_q$ code. 
\end{corollary}
\begin{proof}
	From Theorem \ref{th.conj111}, we take $\C_1$ as a $1$-MDS $[q-1,k_1,q-k_1-1]_q$ code 
	and $\C_2$ be a $1$-MDS $[q-1,k_2,q-k_2-1]_q$ code. 
	Then $\C_2^{\perp}$ is a $1$-MDS $[q-1,q-k_2-1,k_2]_q$ code. 
	By considering the code $\PP(\C_2^{\perp},\C_1)$, 
	the result follows from Theorem \ref{th.m-MDS codes from (u,u+v) construction} 1). 
\end{proof}

We give a specific example to illustrate how Theorem \ref{th.m-MDS codes from (u,u+v) construction} works.

\begin{example}\label{exam.(u,u+v)} Note that we denote the resulting optimal codes in bold and mark derived almost optimal codes with a $^\star$.
	Using the  current $\textsc{Magma}$ $\texttt{BKLC}$ database \cite{magma,codetable}, there is a $1$-MDS $\bf[6,2,4]_2$ code $\C_1$ 
	and its dual is a $1$-MDS $\bf[6,4,2]_2$ code $\C_2$. 
	Applying Lemma \ref{lem.m-MDS codes via PS codes}, one can get more $1$-MDS codes, such as   
	the $1$-MDS $[5,4,1]_2^\star$ code $\C_3$ and $1$-MDS $\bf[5,3,2]_2$ code $\C_4$. 
	Then $\C_5=\C_3^{\perp}$ is a $1$-MDS $[5,1,4]_2^\star$ code and $\C_6=\C_4^{\perp}$ is a $1$-MDS $\bf[5,2,3]_2$ code. 
	By Theorems \ref{th.m-MDS codes from (u,u+v) construction}, we further derive some $\ell$-MDS codes in Table \ref{tab:2}.   
	In addition, one can note that more $\ell$-MDS codes can be obtained by repeatedly applying 
	Theorem \ref{th.m-MDS codes from (u,u+v) construction} to these $\ell$-MDS codes listed in Table \ref{tab:2}.
	\begin{table}[h]
		%\footnotesize
		\centering
		\caption{Some $\ell$-MDS codes from Theorem \ref{th.m-MDS codes from (u,u+v) construction}}
		\label{tab:2}       % Give a unique label
		%\resizebox{\textwidth}{!}{
			\begin{tabular}{c|c|c|c|l}
				\hline
				$\C$ &  $\D$ & $\PP(\C,\D)$ & $\ell$-MDS property & Reference \\ \hline \hline 
				$\C_1$ & $\C_2$ & $\bf[12,6,4]_2$ & $3$-MDS & Theorem \ref{th.m-MDS codes from (u,u+v) construction} 1), 2) or 3) \\ 
				$\C_3$ & $\C_5$ & $[10,5,2]_2$ & $4$-MDS & Theorem \ref{th.m-MDS codes from (u,u+v) construction} 1) \\ 
				$\C_3$ & $\C_6$ & $[10,6,2]_2^\star$ & $3$-MDS & Theorem \ref{th.m-MDS codes from (u,u+v) construction} 1) or 2) \\ 
				
				$\C_4$ & $\C_5$ & $\bf[10,4,4]_2$ & $3$-MDS & Theorem \ref{th.m-MDS codes from (u,u+v) construction} 1) or 3) \\ 
				$\C_4$ & $\C_6$ & $[10,5,3]_2^\star$ & $3$-MDS & Theorem \ref{th.m-MDS codes from (u,u+v) construction} 4) \\ 
				$\C_2$ & $\C_1$ & $[12,6,2]_2$ & $5$-MDS & Theorem \ref{th.m-MDS codes from (u,u+v) construction} 4) \\ 
				$\C_6$ & $\C_4$ & $[10,5,2]_2$ & $4$-MDS & Theorem \ref{th.m-MDS codes from (u,u+v) construction} 4) \\ 
				\hline
			\end{tabular}
			%}
	\end{table}
	%\begin{itemize}
	%    \item From Theorem \ref{th.m-MDS codes from (u,u+v) construction} 1), 2) or 3), $\PP(\C_3,\C_5)$ is a $3$-MDS $\bf[12,6,4]_2$ code; 
	%    \item From Theorem \ref{th.m-MDS codes from (u,u+v) construction} 1), $\PP(\C_5,\C_7)$ is a $4$-MDS $[10,5,2]_2$ code; 
	%    \item From Theorem \ref{th.m-MDS codes from (u,u+v) construction} 1) or 2), $\PP(\C_5,\C_8)$ is a $3$-MDS $[10,6,2]_2^\star$ code; 
	%    \item From Theorem \ref{th.m-MDS codes from (u,u+v) construction} 1) or 3), $\PP(\C_6,\C_7)$ is a $3$-MDS $\bf[10,4,4]_2$ code; 
	%    \item From Theorem \ref{th.m-MDS codes from (u,u+v) construction} 4), $\PP(\C_6,\C_8)$ is a $3$-MDS $[10,5,3]_2^\star$ code; 
	%    \item From Theorem \ref{th.m-MDS codes from (u,u+v) construction} 4), $\PP(\C_4,\C_3)$ is a $5$-MDS $[12,6,2]_2$ code; 
	%    \item From Theorem \ref{th.m-MDS codes from (u,u+v) construction} 4), $\PP(\C_8,\C_6)$ is a $4$-MDS $[10,5,2]_2$ code.  
	%\end{itemize}
\end{example}

\section{General results of $\ell$-MDS codes}\label{sec5} 

In this section, we focus on some general results of $\ell$-MDS codes involving their characterization, weight distributions and bounds. 
Some of them have been studied in the literature for $1$-MDS, $2$-MDS, or $3$-MDS codes. One can note that our results generalize, simplify or improve them.

\subsection{Characterization and weight distributions of $\ell$-MDS codes}

In this subsection, we give an unified characterization and weight distribution formula of $\ell$-MDS codes. 
%One can notice that these general results generalize or improve some known conclusions in the literature. 

\begin{theorem}\label{th.d(C)+d(DC)=n-2m+2}
	Let $\C$ be an $[n,k]_q$ linear code and $\ell\geq 1$ be an integer. 
	If $n>(\ell-1)q+\ell+k-2$ and $k>(\ell-1)q+\ell-2$, 
	then $\C$ is an $\ell$-MDS code if and only if $d(\C)+d(\C^{\perp})=n-2\ell+2$.         
\end{theorem}
\begin{proof}
	Suppose that $\C$ is an $[n,k]_q$ linear code. By Definition \ref{def.m-MDS} and Lemma \ref{prop.m-MDS}, 
	if $\C$ is $\ell$-MDS, we have $d(\C)=n-k-\ell+1$ and $d(\C^{\perp})=k-\ell+1$. Hence, $d(\C)+d(\C^{\perp})=n-2\ell+2$. 
	
	Conversely, on one hand, it follows from Lemma \ref{prop.GHW} 1) that 
	$$\max\{n+1-d_r(\C^{\perp})\mid r=1,2,\ldots,n-k\}=n+1-d(\C^{\perp}).$$ 
	Since $d(\C)+d(\C^{\perp})=n-2\ell+2$, we have $n+1-d(\C^{\perp})=d(\C)+2\ell-1$. 
	With Lemmas \ref{prop.GHW} 1) and 2), we have $$\max\{n+1-d_r(\C^{\perp})\mid r=1,2,\ldots,n-k\}\geq n-k.$$    
	Thus, we have that $d(\C)\geq n-k-2\ell+1$. 
	On the other hand, it follows from Lemmas \ref{prop.GHW} 1) and 3) that 
	$\min\{d_r(\C)\mid r=1,2,\ldots,k\}=d(\C)\leq n-k+1$. 
	Hence, we conclude that $$n-k-2\ell+1\leq d(\C)\leq n-k+1.$$ 
	
	Next, we prove that $d(\C)=n-k-\ell+1$ and hence $d(\C^{\perp})=k-\ell+1$. 
	It then follows that $\C$ is $\ell$-MDS. 
	To this end, we consider the following three cases. 
	
	\begin{enumerate}
		\item[]
	{\textbf{Case 1.-}} $d(\C)\in \{n-k+1, n-k-2\ell+1\}$. If $d(\C)=n-k+1$, then $\C$ is MDS and so does $\C^{\perp}$. 
	It follows that $d(\C)+d(\C^{\perp})=n+2.$ 
	This contradicts to the fact that $d(\C)+d(\C^{\perp})=n-2\ell+2$ and $\ell\geq 1$. 
	If $d(\C)=n-k-2\ell+1$, then $d(\C^{\perp})=k-\ell+1$ and hence $\C^{\perp}$ is MDS. 
	Similarly, this also yields a contradiction.  
	
	\item[] {\textbf{Case 2.-}} $d(\C)\in \{n-k, n-k-1, \ldots, n-k-\ell+2\}$. 
	We take $d(\C)=n-k-\ell+2$ as an example and note that other subcases are similar. 
	Since $n>(\ell-1)q+\ell+k-2$ and $k>(\ell-1)q+\ell-2>(\ell-2)q+\ell-3$, it turns out from Lemma 
	\ref{lem.m-MDS sufficient condition} that the $[n,k,n-k-\ell+2]_q$ linear code $\C$ 
	is an $(\ell-1)$-MDS code. Clearly, it follows that $d(\C)+d(\C^{\perp})=n-2\ell+4$, which is a contradiction.  
	%This also yields a contradiction.  
	
	\item[] {\textbf{Case 3.-}} $d(\C)\in \{n-k-\ell, n-k-\ell-1, \ldots, n-k-2\ell+2\}$.  
	Here, we take $d(\C)=n-k-\ell$ as an example and note that other subcases are similar. 
	If $d(\C)=n-k-\ell$, then $d(\C^{\perp})=k-\ell+2$. 
	%Since $n>(\ell-1)q+\ell+k-2$ and $k>(\ell-1)q+\ell-2$, we get 
	%$k>(\ell-1)q+\ell-2$ and $n>(\ell-1)q+\ell+k-2>(\ell-2)q+\ell+k-3$. 
	%According to Lemma \ref{lem.m-MDS sufficient condition}, 
	%the $[n,n-k,k-\ell+2]_q$ linear code $\C^{\perp}$ is an $(\ell-1)$-MDS code. 
	Since $n>(\ell-1)q+\ell+k-2$ and $k>(\ell-1)q+\ell-2$, we get 
	$n>(\ell-1)q+\ell+n-k-2$ and $n-k>(\ell-1)q+\ell-2>(\ell-2)q+\ell-3$. 
	According to Lemma \ref{lem.m-MDS sufficient condition}, 
	the $[n,n-k,k-\ell+2]_q$ linear code $\C^{\perp}$ is an $(\ell-1)$-MDS code. 
	%It deduces once again that $d(\C)+d(\C^{\perp})=n-2m+4.$ 
	This is an impossible fact. 
\end{enumerate}
	To summarize $\textbf{Cases 1-3}$ above, we conclude that $d(\C)\notin \{n-k-2\ell+1,n-k-2\ell+2,\ldots,n-k-\ell,n-k-\ell+2,n-k-\ell+3, \ldots, n-k+1\}$. 
	It then follows that $d(\C)=n-k-\ell+1$ and $d(\C^{\perp})=k-\ell+1$. 
	Hence, $\C$ is $\ell$-MDS, and we have  completed the proof. 
\end{proof}

Faldum $et~al.$ \cite[Theorem 9]{FW1997-sAMDS} theoretically determined the weight distribution formula for an $[n,k,n-k-\ell+1]_q$ code as follows:   
\begin{align}\label{eq.weight distribution known}
	\small
	\begin{split}
		A_{n-k+\ell+s} = & \binom{n}{k-\ell-s}\sum_{j=0}^{s}(-1)^j\binom{n-k+\ell+s}{j}(q^{\ell+s-j}-1)  \\
		& +\sum_{i=n-k-\ell+1}^{n-k+\ell-1}\sum_{t=k-\ell+1}^{n-i}\left[(-1)^{t-k+\ell+s}\binom{n-i}{k-\ell-s}\binom{n-i-k+\ell+s}{n-i-t}\right]A_i,
	\end{split}
\end{align}
where $\ell\geq 1$ and $0\leq s\leq k-\ell$. 
Note that the weight distribution formula described in Equation (\ref{eq.weight distribution known}) involves many complex summation terms. 
In the following, we simplify the weight distribution formula via some properties of combinatorial numbers.  
%which has significantly less summation terms when $m\geq 2$. %To simplify the weight distribution formula, we use a math

\begin{theorem}\label{th.weight distribution}
	Let $\C$ be an $\ell$-MDS $[n,k]_q$ code, where $\ell\geq 1$. If $s\in \{0,1,\ldots,k-\ell\}$, then  
	\begin{align}\label{eq.A_{n-k+m+s}}
		\begin{split}
			A_{n-k+\ell+s} = & \binom{n}{k-\ell-s}\sum_{j=0}^{s}(-1)^j\binom{n-k+\ell+s}{j}(q^{\ell+s-j}-1)  \\
			& + \sum_{i=n-k-\ell+1}^{n-k+\ell-1}(-1)^{s+1}\binom{n-i}{k-\ell-s}\binom{n-i-k+\ell+s-1}{s}A_i.     
		\end{split}
	\end{align}
	If $s\in \{0,1,\ldots,n-k-\ell\}$, then 
	\begin{align}\label{eq.A'_{k+m+s}}
		\begin{split}
			A^{\perp}_{k+\ell+s} = & \binom{n}{k+\ell+s}\sum_{j=0}^{s}(-1)^j\binom{k+\ell+s}{j}(q^{\ell+s-j}-1)  \\
			& + \sum_{i=k-\ell+1}^{k+\ell-1}(-1)^{s+1}\binom{n-i}{k+\ell+s-i}\binom{k+\ell+s-i-1}{s}A^{\perp}_i.     
		\end{split}
	\end{align}
\end{theorem}
\begin{proof}
	We have that 
	\begin{align*}
		\begin{split}
		 \sum_{t=k-\ell+1}^{n-i}(-1)^{t-k+\ell+s}\binom{n-i-k+\ell+s}{n-i-t}  
			= & (-1)^{s+1}\frac{s+1}{n-i-k+\ell+s}\binom{n-i-k+\ell+s}{n-i-k+\ell-1} \\ 
			= & (-1)^{s+1}\frac{s+1}{n-i-k+\ell+s}\binom{n-i-k+\ell+s}{s+1} \\ 
			= & (-1)^{s+1}\binom{n-i-k+\ell+s-1}{s}.
		\end{split}
	\end{align*}
	It follows from Equation (\ref{eq.weight distribution known}) that Equation (\ref{eq.A_{n-k+m+s}}) holds. 
	Moreover, one can get Equation (\ref{eq.A'_{k+m+s}}) from Equation (\ref{eq.A_{n-k+m+s}}) by duality. 
	This completes the proof. 
\end{proof}

%Furthermore, we have the following bounds on numbers of minimum weight codewords of an $m$-MDS code and its dual code. 
Based on the simplified formulas in Theorem \ref{th.weight distribution}, we  have two bounds on the number of 
minimum weight codewords of an $\ell$-MDS code and its dual code.

\begin{corollary}\label{coro.A and A' minimum weight}
	Let $\C$ be an $\ell$-MDS $[n,k]_q$ code, where $\ell\geq 1$. Then  
	\begin{align}\label{eq.A minimun}
		A_{n-k-\ell+1}\leq \left\lfloor \frac{\binom{n}{k-\ell}(q^\ell-1)}{\binom{k+\ell-1}{k-\ell}} \right\rfloor  
	\end{align}  
	with equality if and only if $A_i=0$ for each $n-k-\ell+2\leq i\leq n-k+\ell$. 
	And 
	\begin{align}\label{eq.A' minimun}
		A^{\perp}_{k-\ell+1}\leq \left\lfloor \frac{\binom{n}{k+\ell}(q^\ell-1)}{\binom{n-k+\ell-1}{2\ell-1}} \right\rfloor 
	\end{align}
	with equality if and only if $A^{\perp}_i=0$ for each $k-\ell+2\leq i\leq k+\ell$. 
\end{corollary}
\begin{proof}
	Taking $s=0$ in Equation (\ref{eq.A_{n-k+m+s}}), one has $$A_{n-k+\ell}=\binom{n}{k-\ell}(q^\ell-1)-\sum_{i=n-k-\ell+1}^{n-k+\ell-1}\binom{n-i}{k-\ell}A_i.$$  
	Thus it follows that $$\binom{k+\ell-1}{k-\ell}A_{n-k-\ell+1}=\binom{n}{k-\ell}(q^\ell-1)-\sum_{i=n-k-\ell+2}^{n-k+\ell}\binom{n-i}{k-\ell}A_i.$$
	Since $A_i$ is a non-negative integer for each $0\leq i\leq n$, the desired Equation (\ref{eq.A minimun}) holds. 
	By duality, Equation (\ref{eq.A' minimun}) follows straightforward from Equation (\ref{eq.A minimun}). 
	Note that conditions under which these two inequalities become equalities are apparent. 
\end{proof}

\begin{remark}\label{rem.mMDS weight distribution}  $\quad$ 
	%Additionally, Theorem \ref{th.weight distribution} yields the same result  with \cite[Theorem 4.1]{DL1995-NMDS} and \cite[Corollary 15]{FW1997-sAMDS} if $m=1$. 
	\begin{enumerate}
		\item   From Theorem \ref{th.d(C)+d(DC)=n-2m+2}, under certain conditions,  
		one can check  that $\C$ is a $1$, $2$, or $3$-MDS code %(i.e., $m=1$, $2$, or $3$, respectively) 
		if and only if $d(\C)+d(\C^{\perp})=n-2$, $n-4$, or $n-6$. 
		These cases are respectively the same with \cite[Corollary 3.3]{DL1995-NMDS}, \cite[Theorem 5]{T2012-NNMDS}, 
		and \cite[Theorem 4.4]{LL2014-m-MDS-Liao}. 
		Hence, Theorem \ref{th.d(C)+d(DC)=n-2m+2} can be seen as a generalization of them.  
		
		\item   Let $\C$ be an $\ell$-MDS $[n,k]_q$ code, where $\ell\geq 1$. 
		On one hand, Corollary \ref{coro.A and A' minimum weight} %implies that $$A_{n-k}\leq \left\lfloor \binom{n}{k-1}\frac{q-1}{k} \right\rfloor\ {\rm and}\ A'_k\leq \left\lfloor\binom{n}{k+1}\frac{q-1}{n-k}\right\rfloor,$$ 
		yields the same result with Lemma \ref{lem.NMDS weight distribution} provided that $\ell=1$.  
		On the other hand, \cite[Theorem 1.1.16]{TV1991-AG codes} states that 
		\begin{align}\label{eq.*}
			A_{n-k-\ell+1} \leq \binom{n}{k+\ell-1}(q-1) 
		\end{align}
		and 
		\begin{align}\label{eq.**}
			A^{\perp}_{k-\ell+1} \leq \binom{n}{k-\ell+1}(q-1).
		\end{align}
		%$A_{n-k-m+1}\leq \binom{n}{k+m-1}(q-1)$ ($*$) and $A^{\perp}_{k-m+1}\leq \binom{n}{k-m+1}(q-1)$ ($**$). 
		Note that 
		\begin{align*}
			\begin{split}
				 \left \lfloor \frac{\binom{n}{k-\ell}(q^\ell-1)}{\binom{k+\ell-1}{k-\ell}} \right \rfloor < \binom{n}{k+\ell-1}(q-1)  
				\Leftrightarrow\ &  \frac{\binom{n}{k-\ell}(q^\ell-1)}{\binom{k+\ell-1}{k-\ell}}<\binom{n}{k+\ell-1}(q-1) \\ 
				\Leftrightarrow\ & \frac{q^\ell-1}{q-1}<\frac{\binom{n}{k+\ell-1} \binom{k+\ell-1}{k-\ell}}{\binom{n}{k-\ell}}=\binom{n-k+\ell}{2\ell-1}. \\
			\end{split}
		\end{align*}
		Hence, Equation (\ref{eq.A minimun}) in Corollary \ref{coro.A and A' minimum weight} gives an improved upper bound  
		with respect to Equation (\ref{eq.*}) if $\frac{q^\ell-1}{q-1}<\binom{n-k+\ell}{2\ell-1}$. 
		Similarly, Equation (\ref{eq.A' minimun}) in Corollary \ref{coro.A and A' minimum weight} also gives an improved upper bound  
		with respect to Equation (\ref{eq.**}) if $\frac{q^\ell-1}{q-1}<\binom{k+\ell}{2\ell-1}$.     
	\end{enumerate}
\end{remark}

\begin{example}\label{exam.[9,5,3]_3 weight distribution} 
	Let $\C$ be a $2$-MDS $[9,5,3]_3$ code with a generator matrix   
	$$G=\left[\begin{array}{c}
		100000112 \\
		010002110 \\
		001001200 \\
		000100021 \\ 
		000011111
	\end{array}\right].$$ Then $\C^{\perp}$ has parameters $[9,4,4]_3$. 
	%Two concrete examples supporting above statements are as follows.  Let $\C_2$ be the $[9,5,3]_3$ linear code generated by $G_2$ and the dual of $\C_2$ is the linear code $\C_2^{\perp}$ with parameters $[9,4,4]_3$. Then $\C_2$ is a $2$-MDS code. 
	On one hand, we have checked with $\textsc{Magma}$ \cite{magma} that  $A_3=10$, $A_4=18$, $A_5=54$, $A^{\perp}_4=14$, $A^{\perp}_5=8$ and $A^{\perp}_6=26$.     
	Hence, it follows from Theorem \ref{th.weight distribution} that $\C$ and $\C^{\perp}$ have the following polynomial weight enumerators  
	$$A(z)=1+10z^3+18z^4+54z^5+76z^6+54z^7+18z^8+12z^9, \, A^{\perp}(z)=1+14z^4+8z^5+26z^6+22z^7+10z^8,$$ respectively. 
	On the other hand, from Corollary \ref{coro.A and A' minimum weight} we have that  $A_3\leq 33$ and $A^{\perp}_4\leq 28$. 
	However, Theorem 1.1.16 in \cite{TV1991-AG codes} yields $A_3\leq 168$ and $A^{\perp}_4\leq 252$. 
	It is clear that Corollary \ref{coro.A and A' minimum weight} gives tighter upper bounds on $A_3$ and $A^{\perp}_4$. 
	%On the other hand, note that $A_3$ and $A^{\perp}_4$ are from weight distributions above actually $10$ and $14$, which indeed satisfy the bounds $A_3\leq 33$ and $A^{\perp}_4\leq 28$.   
\end{example}

\subsection{Bounds on maximum lengths of $\ell$-MDS codes} 

%By $N^{m}(k, q)$ we denote the maximum possible length $n$, for which there exists an $[n, k,n-k-m+1]_q$ code and $N_{m}(k, q)$ we denote the maximum possible length $n$, for which there exists an $[n, k]_q$ $m$-MDS code. 
%In other words, 
Let 
\begin{align}
	N^{\ell}(k,q)=\max\{n\mid \text{there\ exists\ an}\ [n,k,n-k-\ell+1]_q\ \text{code}\}
\end{align}
and 
\begin{align}
	N_{\ell}(k,q)=\max\{n\mid \text{there\ exists\ an}\ [n,k]_q\ \ell\text{-MDS\ code}\}. 
\end{align}
Then we have the following results.

\begin{theorem}\label{th.m-mds bounds}
  The following statements hold. 
	\begin{enumerate}
		\item   $N_\ell(k,q)\leq N^{\ell}(k,q)$.
		\item   If $N^\ell(k,q)>\ell q+\ell+k-1$ and $k>(\ell-1)q+\ell-2$, then $N_\ell(k,q)=N^{\ell}(k,q)$. 
		\item   If $k\geq 2$, then $N^\ell(k,q)\leq (\ell+1)q+\ell+k-1$. 
		\item  If $\ell\geq 1$, then $N^\ell(k,q)\geq N^{\ell-1}(k,q)+1$. 
		\item   If $k>(\ell+1)q+\ell-1$ and $\ell\geq 1$, then $N^{\ell}(k,q)\leq \ell q+\ell+k-1$. 
		\item   For any $1\leq s\leq k$, if $N_\ell(k-s+1,q)>(\ell-1)q+k+\ell-s-1$ and $k>(\ell-1)q+\ell+s-3$, then $N_\ell(k,q)\leq N_{\ell}(k-s,q)+s$. 
	\end{enumerate}
\end{theorem}
\begin{proof}$\quad$
	
	\begin{enumerate}
	\item[1)] By definition, the result  is obvious. 
	
	\item[2)] The desired result follows from Lemma \ref{lem.m-MDS sufficient condition}. 
	
	\item[3-5)] These results have been shown in Theorems 8 and 9 1) as well as 9 2) of \cite{TCY2014-m-MDS-Tong}. 
	\item[6)] Since $N_\ell(k-s+1,q)>(\ell-1)q+k+\ell-s-1$ and $k>(\ell-1)q+\ell+s-3$, it turns out from Lemma \ref{lem.m-MDS codes via PS codes} 2)  
	that an $[N_\ell(k,q)-s,k-s]_q$ $\ell$-MDS code exists if there is an $[N_\ell(k,q),k]_q$ $\ell$-MDS code. 
	By definition, we further have $N_\ell(k-s,q)\geq N_\ell(k,q)-s$, i.e., $N_\ell(k,q)\leq N_\ell(k-s,q)+s$. 
		\end{enumerate} 
	%For the proof of 6), we note that the existence of an $m$-MDS $[n,k]_q$ code implies the existence of an 
\end{proof}

%Obviously, $N_m(k,q)\leq N^{m}(k,q)$ and 
%$N_m(k,q)=N^{m}(k,q)$ if $N^m(k,q)>mq+m+k-1$ and $k>(m-1)q+m-2$ by 
%Lemma \ref{lem.m-MDS sufficient condition}. 

In fact, we can also derive an improved upper bound for $N_\ell(k,q)$ with respect to known bounds presented in 
Theorems \ref{th.m-mds bounds} 3) and 5) when $q=2$ and $k\geq 3$.  

\begin{theorem}\label{th.bounds for q=2}
	   If $k\geq 3$, then 
	\begin{align}\label{eq.bounds for q=2}
		N(k,2)_\ell\leq N(k,2)^\ell\leq k+2\ell+\left\lfloor \frac{\ell}{3} \right\rfloor+1. 
	\end{align}
\end{theorem}
\begin{proof}
	Suppose that there exists an $[n,k,n-k-\ell+1]_2$ linear code and $n-k-\ell+1\equiv b~({\rm mod}~4)$, where $b=0,1,2,3$. 
	If $k\geq 3$, it follows from the well-known Griesmer bound \cite{Griesmer} that 
	\begin{align*}
		\begin{split}
			n & \geq \sum_{i=0}^{k-1}\left\lceil \frac{n-k-\ell+1}{2^i} \right\rceil \\ 
			& \geq n-k-\ell+1+\left\lceil \frac{n-k-\ell+1}{2} \right\rceil+\left\lceil \frac{n-k-\ell+1}{4} \right\rceil+k-3. 
		\end{split}
	\end{align*}
	Thus, we have that $\ell+2\geq \left\lceil \frac{n-k-\ell+1}{2} \right\rceil+\left\lceil \frac{n-k-\ell+1}{4} \right\rceil$. 
	Write $n-k-\ell+1=4a+b$, where $a\geq 0$ and $0\leq b< 4$ are two integers. Then 
	\begin{align*}
		\left\lceil \frac{n-k-\ell+1}{2} \right\rceil=2a+\left\lceil \frac{b}{2} \right\rceil=\left\{\begin{array}{ll}
			2a, & {\rm if}\ b=0,\\
			2a+1, & {\rm if}\ b=1,2, \\
			2a+2, & {\rm if}\ b=3
		\end{array}\right.
	\end{align*}
	and 
	\begin{align*}
		\left\lceil \frac{n-k-\ell+1}{4} \right\rceil=a+\left\lceil \frac{b}{4} \right\rceil=\left\{\begin{array}{ll} 
			a, & {\rm if}\ b=0,\\
			a+1, & {\rm if}\ b=1,2,3.
		\end{array}\right.
	\end{align*}
	We have the following three cases. 
	\begin{description}
		\item[] {\bf Case 1}.- If $b=0$,  we have $\ell+2\geq 3a$, i.e., $a\leq \frac{\ell+2}{3}$. Hence, one has  
	\begin{align*}
		\left\{\begin{array}{l}
			\left\lceil \frac{n-k-\ell+1}{2} \right\rceil\leq \frac{2\ell+4}{3},\\ \\
			\left\lceil \frac{n-k-\ell+1}{4} \right\rceil\leq \frac{\ell+2}{3}. 
		\end{array}\right.
	\end{align*}
	It implies that $n\leq k+2\ell+\left\lfloor \frac{\ell+2}{3} \right\rfloor+1$. 
	
		\item[] {\bf Case 2}.- If $b=1,2$, we have $\ell+2\geq 3a+2$, i.e., $a\leq \frac{\ell}{3}$. Therefore
	\begin{align*}
		\left\{\begin{array}{l}
			\left\lceil \frac{n-k-\ell+1}{2} \right\rceil\leq \frac{2\ell+3}{3},\\  \\ 
			\left\lceil \frac{n-k-\ell+1}{4} \right\rceil\leq \frac{\ell+3}{3}.
		\end{array}\right.
	\end{align*}
	It implies that $n\leq k+2\ell+\left\lfloor \frac{\ell}{3} \right\rfloor+1$. 
	
		\item[] {\bf Case 3}.- If $b=3$,  we have $\ell+2\geq 3a+3$, i.e., $a\leq \frac{\ell-1}{3}$. Thus,  
	\begin{align*}
		\left\{\begin{array}{l}
			\left\lceil \frac{n-k-\ell+1}{2} \right\rceil\leq \frac{2\ell+4}{3},\\  \\ 
			\left\lceil \frac{n-k-\ell+1}{4} \right\rceil\leq \frac{\ell+2}{3}. 
		\end{array}\right.
	\end{align*}
	As in $\textbf{Case 1}$, it implies that $n\leq k+2\ell+\left\lfloor \frac{\ell+2}{3} \right\rfloor+1$. 
\end{description}	
	In summary, the desired result follows from Theorem \ref{th.m-mds bounds} 1). 
\end{proof}

\begin{remark} 
	By a tedious but direct computation, one can deduce that Theorem \ref{th.bounds for q=2} provides a tighter upper bound 
	than Theorem \ref{th.m-mds bounds} 3) if $\ell\geq 2$ and Theorem \ref{th.m-mds bounds} 5) if $\ell\geq 5$. 
	As a application, Theorem \ref{th.bounds for q=2} is more effective in determining the existence of binary $\ell$-MDS codes than 
	Theorems \ref{th.m-mds bounds} 3) and 5). For example, it is easily deduced from Theorem \ref{th.bounds for q=2} that 
	$\ell_1$-MDS $[959,312]_2$ codes and $\ell_2$-MDS $[683,495]_2$ codes do not exist for $216\leq \ell_1 \leq 362$ and $63\leq \ell_2 \leq 80$. 
	However, these results can not be obtained from Theorems \ref{th.m-mds bounds} 3) or 5).  
\end{remark}

%\subsection{FSD codes}

\section{Concluding remarks and further research}\label{sec6}

In this paper, we have studied $\ell$-MDS codes. 
First, we have focused on a conjecture on $1$-MDS codes proposed by Heng $et~al.$ \cite{HW2023Conj} 
and completely solve it based on some useful results involves subset sum problems. 
Consequently, infinite families of $1$-MDS codes with general dimensions support $2$-designs are obtained and they are also proved to be optimal locally recoverable codes. 
Then, we have constructed general $\ell$-MDS codes from old ones by using extended codes, expurgated codes, and the $(\uuu,\uuu+\vvv)$ construction. 
As a result, we derived new binary optimal $\ell$-MDS codes and $q$-ary infinite families of $\ell$-MDS (FSD) codes with length $n>q$.    
Finally, we have presented some general results of $\ell$-MDS codes, which generalize, simplify, or improve known conclusions on $1$-MDS, $2$-MDS, and $3$-MDS codes 
in \cite{LL2014-m-MDS-Liao,DL1995-NMDS,T2012-NNMDS,TCY2014-m-MDS-Tong}. 

As a  future research topic, it would be interesting to construct more infinite families of general $\ell$-MDS codes with length $n>q$. 
Another interesting direction is to explore more possible applications of $\ell$-MDS codes, 
such as error-correcting pairs \cite{HL1,HL2}. %and coded caching schemes \cite{TC2016}.   

\end{document}